\documentclass[11pt,a4paper,twoside]{article}
\usepackage{a4wide}
\usepackage[english]{babel}

\newcommand{\color}[1]{}

\usepackage{amsmath,amssymb,amsfonts,amscd}
\usepackage{dsfont}
\usepackage[mathscr]{eucal}

\newcommand {\anumber}  {}    
\newcommand {\eqip} [1] {#1}  

\newcommand {\eps} {\varepsilon}

\newcommand {\nosum} {\qquad \mbox{\scriptsize (no sum over the indices)}}

\newcommand {\half} [1][1] {\ensuremath{\frac{#1}{2}}}

\newcommand {\ket} [1] {\ensuremath{\left| \!\!
                                      \begin{array}{c}
                                        #1 \\
                                      \end{array} \!\!
                                    \right>}}

\newcommand {\sups} [1] {\ensuremath{^{\textrm{\scriptsize #1}}}}

\newcommand {\mycite} [1] {\sups{\color{magenta}\cite{#1}}}

\DeclareMathOperator{\ad}{ad}

\renewcommand{\d}{\mbox{d}}

\newcommand {\ZZ}{\ensuremath{\mathds{Z}}}

\newcommand {\RR}{\ensuremath{\mathds{R}}}
\newcommand {\CC}{\ensuremath{\mathds{C}}}

\newcommand {\unit}{\ensuremath{\mathds{1}}}

\newcommand {\topk}{\vec{k}}
\newcommand {\Bbk} {b(\topk)}
\newcommand {\sogen}{Y}

\newcommand {\Egamma} {\ensuremath{\overline{\gamma}}}

\newcommand {\ICS} {\ensuremath{\mathcal{I}}}
\newcommand {\JCS} {\ensuremath{\mathcal{J}}}
\newcommand {\KCS} {\ensuremath{\mathcal{K}}}

\newcommand {\qi} {\imath}
\newcommand {\qj} {\jmath}
\newcommand {\qk} {\kappa}

\newcommand {\qBos} {\mathsf{w}}
\newcommand {\qq}   {\mathsf{e}}
\newcommand {\qD}   {\mathsf{D}}

\newcommand {\Jang} {\ensuremath{J}} 
\newcommand {\MS} {\ensuremath{\mathcal{M}}}
\newcommand {\pauli} {\ensuremath{\sigma}}

\newcommand {\LieG} {\ensuremath{\mathfrak{g}}}

\newcommand {\ofr} {\ensuremath{e}}
\newcommand {\ofc} {\ensuremath{\theta}}

\newcommand {\slashed} [1][0ex] {\hspace{-1.1ex}\hspace{#1}/\hspace{-.06ex}\hspace{-#1}}
\newcommand {\Slashed} {\slashed[-.35ex]}
\newcommand {\slD}     {D\Slashed}

\newcommand {\cGMpm} {+} 
\newcommand {\cGMmp} {-} 

\begin{document}

\pagenumbering{arabic}

\title{  \bf Supersymmetric Quantum Mechanics of Magnetic Monopoles: A Case Study         }
\author{                                                                                \\
             Erik Jan de Vries\footnote{erik@ma.hw.ac.uk}~ and Bernd J. Schroers\footnote{bernd@ma.hw.ac.uk}                                       \\
                                                                                        \\
         \it Department of Mathematics and Maxwell Institute for Mathematical Sciences, \\
         \it Heriot-Watt University, Edinburgh, EH14 4AS, UK                                \\
                                                                                        \\}
\date{      February 3 , 2009                                                         \\
             ~                                                                          \\
             \textsf{\small{EMPG-08-18}}                                                \\}

\maketitle
\thispagestyle{empty}

\begin{abstract}
We study, in detail, the supersymmetric quantum mechanics of charge-$(1,1)$ monopoles in $N=2$ supersymmetric Yang-Mills-Higgs theory with gauge group $SU(3)$ spontaneously broken to $U(1)\times U(1)$. We use the moduli space approximation of the quantised dynamics, which can be expressed in two equivalent formalisms: either one describes quantum states by Dirac spinors on the moduli space, in which case the Hamiltonian is the square of the Dirac operator, or one works with anti-holomorphic forms on the moduli space, in which case the Hamiltonian is the Laplacian acting on forms. We review the derivation of both formalisms, explicitly exhibit their equivalence and derive general expressions for the supercharges as differential operators in both formalisms. We propose a general expression for the total angular momentum operator as a differential operator, and check its commutation relations with the supercharges. Using the known metric structure of the moduli space of charge-$(1,1)$ monopoles we show that there are no quantum bound states of such monopoles in the moduli space approximation. We exhibit scattering states and compute corresponding differential cross sections.
\end{abstract}



\section{Introduction}
One of the most intriguing aspects of supersymmetric Yang-Mills-Higgs theory is that it generically contains magnetic monopoles as classical, solitonic solutions with properties which appear to be dual to those of the perturbative electrically charged quantum particles in the theory. The strongest formulation of this duality is the electric-magnetic duality conjecture of Montonen and Olive\mycite{MontOlive:MMGP}, according to which the physics of massive electrically charged particles (W-bosons) in the theory should be equivalent to the physics of magnetically charged particles (magnetic monopoles) in the dual theory, where the gauge group is replaced by its dual and the coupling constant inverted. If the conjecture is correct it should be possible to investigate the physics of strongly coupled electric particles by studying the physics of magnetic monopoles at weak coupling, using perturbative or semiclassical techniques.

The properties of monopoles in supersymmetric Yang-Mills-Higgs theory have been studied extensively, often motivated by the Montonen-Olive duality conjecture or its generalisation to S-duality\mycite{Sen:DMBS}. A crucial tool in these studies has been the moduli space approximation\mycite{Manton:ScatteringBPSMon} of quantum dynamics of supersymmetric monopoles. Much is known about this approximation, but previous papers have described rather general aspects (like the Hilbert space and Hamiltonian of the theory) or have focused on the calculation of lowest bound states, the so-called BPS states, which play a key role in testing the S-duality conjecture. In this paper we go further, looking for higher bound states, studying scattering states and investigating other observables like angular momentum and supercharges in the theory.

Our goal is to illustrate how one can use semiclassical techniques to compute details of quantised interactive monopole dynamics, with a view to probing strongly coupled physics of electric particles via duality.
Even though the Montonen-Olive conjecture can only possibly hold in $N=4$ supersymmetric Yang-Mills-Higgs theory, we study the $N=2$ case in this paper, which should be regarded as a stepping stone towards the study of the $N=4$ case in a future publication.
We focus on Yang-Mills-Higgs theory with gauge group $SU(3)$ broken maximally to $U(1)\times U(1)$, and investigate the quantised dynamics of two distinct monopoles in this theory, using moduli space techniques.  We are able to show that there are no bound states in this model, and that the scattering cross section of two distinct monopoles in this theory is the same, at low energies, as that of two BPS monopoles in  the $SU(2)$ theory with the symmetry breaking to $U(1)$.

A secondary purpose of our paper is to exhibit some of the interesting geometrical features of the moduli space approximation to supersymmetric quantum dynamics of monopoles. As we shall explain, quantum states of monopoles in $N=2$ supersymmetric Yang-Mills-Higgs theory can be described in terms of either anti-holomorphic forms or spinors on the moduli space. The equivalence of the two descriptions follows from the hyperk\"ahler property of monopole moduli spaces. In the second viewpoint, both fermionic and bosonic degrees of freedom in the original field theory are encoded in spinors on the moduli space.
This may seem puzzling at first, and this viewpoint is indeed less convenient when it comes to interpreting the moduli space quantum mechanics in terms of the original fields. However, the spinorial viewpoint is interesting not least because of the links with the large literature on spectral properties of Dirac operators which it provides. We will see and exploit this explicitly  in our case study where   earlier work by Comtet and Horv\'athy\mycite{ComHor:DiracTaubNUT} on the Dirac operator on the Taub-NUT manifold provides useful guidance.

In order to summarise our paper in more precise terms we need to review some aspects of the moduli space approximation to quantum dynamics of supersymmetric monopoles. The moduli space approximation was originally introduced by Manton in order to study the classical dynamics of several interacting monopoles\mycite{Manton:ScatteringBPSMon}. The quantisation of the moduli space approximation was first considered in the context of the non-supersymmetric bosonic theory by supposing that quantum states are scalar functions on the moduli space and that the Hamiltonian is proportional to the Laplacian on the moduli space\mycite{GibbonsManton:CQDynBPSMon}. This model was then used\mycite{Manton:MonSkyBS,Schroers:QSBPSMon} to compute bound state energies and scattering cross sections for elastic and inelastic monopole-monopole scattering in Yang-Mills-Higgs theory with gauge group $SU(2)$ broken to $U(1)$.

Gauntlett\mycite{Gauntlett:LowEnDymN=2Mon} subsequently showed that the moduli space approximation to quantum dynamics of monopoles in the $N=2$ supersymmetric extension of the bosonic theory leads to a model where the quantum states are anti-holomorphic forms on the moduli space, and the Hamiltonian is the Laplacian acting on forms. The appropriate model for $N=4$ supersymmetric monopoles has wavefunctions which are arbitrary differential forms on the moduli space and the Hamiltonian is again the Laplacian\mycite{Gauntlett:LowEnDymN=2Mon,Blum:SusyQMMonN=4,Witten:ConstrSusyBr}. This model played a crucial role in the genesis of the S-duality conjecture\mycite{Sen:DMBS}.

Our study of supersymmetric monopole dynamics is organised as follows.
In sections 2 and 3 we fix our conventions and notation, and review relevant aspects of $N=2$ supersymmetric Yang-Mills-Higgs theory and of monopole moduli spaces, with a particular emphasis on their hyperk\"ahler structure.
In section~4 we explain how the quantisation of the moduli space approximation leads to a model where quantum states are spinors on the moduli space. This quantum mechanical model has $N=4$ supersymmetry, and we derive expressions for the four supercharges as (twisted) Dirac operators.
In section~5 we review an alternative quantisation, where the Hilbert space consists of anti-holomorphic forms with respect to one of the complex structures. Again we derive expressions for the four supercharges, this time as (twisted) Dolbeault operators.
Section 6 is devoted to a particularly important observable, namely the angular momentum operator. The naive guess that its components are simply the infinitesimal generators of the $SO(3)$ action on the moduli space cannot be correct since this action does not respect the complex structure used in defining the quantum states. We propose a modification which does act on the Hilbert space of anti-holomorphic forms, and show that it has the required commutation relations with the supercharges.
In section~7 we illustrate the results of the previous sections in a simple example, namely the moduli space of a single monopole. Then, in section~8, we consider the moduli space quantisation of
two distinct monopoles in $N=2$ supersymmetric Yang-Mills-Higgs theory with gauge group $SU(3)$ broken to $U(1)\times U(1)$. The relevant moduli space is an eight-dimensional hyperk\"ahler
manifold of the form $\RR^3\times(\RR\times \MS_{TN})/{\ZZ}$, where $\MS_{TN}$ is the complete self-dual Taub-NUT space with positive mass parameter\mycite{GauLowe:DyonsSDuality,LeeWeinbergYi:EMDSU3Mon,Connell:dyn11mon}. The reason for choosing this example is that the scalar Laplace equation on this space can be solved exactly, and that the complex structures are known. As a result, we are able to exhibit many features of the supersymmetric quantum mechanics explicitly: we show that there are no bound states, and give explicit formulae for differential cross sections.  Finally, section~9 contains an outlook.


\section{Monopole moduli spaces}
We start with a brief review of the construction of monopole moduli spaces. We discuss the field theoretical model that gives rise to monopoles and the interpretation of monopoles as translation invariant instantons in Euclidean $\RR^4$. In the next section we construct the moduli space of BPS monopoles and discuss its hyperk\"ahler structure. We introduce a description of zero-modes in terms of quaternions that will be useful in further calculations.

\subsection{BPS monopoles}
Monopoles appear in certain classes of Yang-Mills-Higgs field theories with spontaneously broken symmetries\mycite{Hooft:Monopole,Polyakov:Monopole}. We study Yang-Mills-Higgs models on space-time with a Lorentzian metric  $\eta$ of signature $(+,-,-,-)$. The Lagrangian that we are interested in is given by
\begin{align}
L \ = \ \int \d^3 x \ \mathcal{L}
 & \ = \ \int \d^3 x \left( - \frac{1}{4} F_{\mu\nu} \cdot F^{\mu\nu} + \half D_{\mu}\Phi \cdot D^{\mu}\Phi \right)\eqip{.} \label{YMHLagrangian}
\end{align}
The fields are in the adjoint representation of the gauge group $G = SU(n)$,  i.e. they take values in the Lie-algebra $\LieG = su(n)$. The dot-product is an invariant inner product on the Lie-algebra.
$F_{\mu\nu} = \partial_{\mu} A_{\nu} - \partial_{\nu} A_{\mu} - e \left[ A_{\mu}, A_{\nu} \right]$ is the field strength, where $-e$ is the coupling constant.
The covariant derivative of the Higgs field $\Phi$ is $D_{\mu}\Phi = \partial_{\mu} \Phi - e \ad A_{\mu} \Phi = \partial_{\mu} \Phi - e \left[ A_{\mu}, \Phi \right]$.
We impose boundary conditions for the fields at infinity to break the symmetry, $\lim_{r\to\infty} \Phi\cdot\Phi = a^2$. We will assume that the symmetry breaking is maximal; specifically, $SU(n)$ is broken to $U(1)^{n-1}$.

The equations of motions are
\begin{align}
D_{\mu} D^{\mu} \Phi & \ = \ 0\eqip{,} &
D_{\mu} F^{\mu\nu} & \ = \ - e \left[ D^{\nu} \Phi , \Phi \right]\eqip{.} \label{eom}
\end{align}
The conjugate momenta to $A_i$ and $\Phi$ are
\begin{align}
E^{i} & \ \equiv \ \frac{\partial \mathcal{L}}{\partial \dot{A}_i} \ = \  F^{i0}\eqip{,} &
\Pi & \ \equiv \ \frac{\partial \mathcal{L}}{\partial \dot{\Phi}} \ = \  D_0 \Phi\eqip{.} \anumber
\end{align}
The conjugate momentum to $A_0$ vanishes, so we must impose Gauss' Law (the equation of motion for $A_0$) as a constraint on the gauge fields. The Lagrangian $L$ and energy $H$ can be written as
\begin{align}
L & \ = \ K - V\eqip{,} &
H & \ = \ K + V\eqip{,} \anumber
\end{align}
where, defining $B_i = \half \epsilon_{ijk} F^{jk}$, the kinetic and potential energy are given by
\begin{align}
K & \ = \ \int \d^3 x \left( \half |E_i|^2 + \half |\Pi|^2 \right)\eqip{,} &
V & \ = \ \int \d^3 x \left( \half |B_i|^2 + \half |D_i \Phi|^2 \right)\eqip{.} \label{KV}
\end{align}

Bogomol'nyi\mycite{Bogomolnyi:Stability} first observed that the potential energy can be written as
\begin{align}
V & \ = \ \int \d^3 x \left( \half \left|B_i \mp D_i \Phi\right|^2 \pm \partial_i \left( \Phi \cdot B_i \right) \right)\eqip{.} \anumber
\end{align}
Then
\begin{align}
\left| \int \d^3 x \partial_i \left( \Phi \cdot B_i \right) \right| & \ = \ \left| \int_{S^2_{\infty}} \d S_i \left( \Phi \cdot B_i \right) \right| \ = \ \frac{4\pi a}{e} \Bbk\eqip{,} \anumber
\end{align}
where $\topk=(k_1,\ldots,k_{n-1})\in\ZZ^{n-1}$ is the topological charge  and $\Bbk$ is a positive, real function of the topological charge which depends on the details of the vacuum expectation value of  the Higgs field.
Hence the potential energy  satisfies the Bogomol'nyi bound
\begin{align}
V & \ \geq \ \frac{4\pi a}{e} \Bbk\eqip{.} \label{BogomolnyiBound}
\end{align}
BPS monopoles have minimal energy, which means that they are static and they saturate the Bogomol'nyi bound. The latter implies that they must satisfy the Bogomol'nyi equations
\begin{align}
B_i & \ = \ \pm D_i \Phi\eqip{.} \label{Bogomolnyi}
\end{align}
Here the upper sign corresponds to monopoles (positive topological charge, i.e. all integers $k_1,\ldots, k_{n-1}$ are $\geq 0$), and the lower sign to anti-monopoles (negative topological charge).
The Bogomol'nyi equations imply the equations of motion for static field configurations.

\subsection{Euclidean 4-space}
\label{E4}
It is convenient to work in the temporal gauge, $A_0 = 0$. We define
\begin{align}
W_i & \ = \ A_i\eqip{,} &
W_4 & \ = \ \Phi\eqip{.} \label{defW}
\end{align}
We can think of $W_{\underline{i}}$ (underlined indices $\underline{i}$ run from 1 to 4) as a connection on Euclidean $\RR^4$, if we introduce a fourth spatial dimension and assume all fields to be independent of this fourth dimension, $\partial_4 \equiv 0$. The covariant derivatives are defined by
\begin{align}
  D_{\underline{i}} & \ = \ \partial_{\underline{i}} \, - e \ad W_{\underline{i}}\eqip{.} \label{defD}
\end{align}
Infinitesimal gauge transformations can be written as $\delta_{\Lambda} W_{\underline{i}} = -\frac{1}{e} D_{\underline{i}} \Lambda$ for some gauge parameter $\Lambda$, which is again taken to be independent of the fourth dimension.

In terms of the field strength $G_{\underline{ij}}$ corresponding to $W_{\underline{i}}$, the equations of motion and Gauss' Law become
\begin{align}
\ddot{W}_{\underline{i}} & \ = \ D_{\underline{j}} G_{\underline{ji}}\eqip{,} &
D_{\underline{i}} \dot{W}_{\underline{i}} & \ = \ 0\eqip{,} \label{GaussLaw}
\end{align}
where dots denote time derivatives. The kinetic and potential energy can be written as
\begin{align}
K & \ = \ \int \d^3 x \ \half \left|\dot{W}_{\underline{i}} \right|^2\eqip{,} &
V & \ = \ \int \d^3 x \ \frac{1}{4} |G_{\underline{ij}}|^2\eqip{.} \label{KV2}
\end{align}
The Bogomol'nyi equations become the (anti-)self-duality equations for the field strength,
\begin{align}
G_{\underline{ij}} & \ = \ \pm \half \eps_{\underline{ijkl}} G_{\underline{kl}}\eqip{.} \anumber
\end{align}
Therefore we may think of monopoles as instantons in Euclidian 4-space that are independent of the fourth dimension.

\subsection{The moduli space approximation}
The moduli space is the space of physically distinct monopoles of minimal energy within a particular topological class $\topk$. Field configurations of monopoles related to each other by gauge transformations are not physically distinct. The moduli space is therefore the space of gauge equivalence classes of BPS monopoles with a particular topological charge. 

We denote the set of finite energy field configurations $W_{\underline{i}}$ by $\mathcal{A}$, and the group of short-range gauge transformations by $\mathcal{G}$. Short-range gauge transformations are those gauge transformations that tend to the identity at infinity. The configuration space is obtained by identifying field configurations that are related via a short-range gauge transformation, $\mathcal{C} = \mathcal{A}/\mathcal{G}$. Long-range gauge transformations, with non-trivial action on the fields at infinity, are excluded from $\mathcal{G}$, because when we allow such gauge transformations to become time dependent, they have a physical effect on the monopoles (turning them into dyons with electric charge\mycite{JulZee:Dyon}), unlike time dependent short-range gauge transformations.

The moduli space for monopoles of charge $\topk$, $\MS_{\topk}$, is the subspace of $\mathcal{C}$ corresponding to static solutions of the Bogomol'nyi equations with topological charge $\topk$.
The idea of the moduli space approximation is to describe the low energy dynamics of monopoles by motion in the moduli space\mycite{Manton:ScatteringBPSMon}. This is an adiabatic description and a  good approximation provided the monopoles move slowly\mycite{MantSam:RadMonSc}. In order
to compute the equation governing the path in the moduli space, we lift it to a path in $\mathcal{A}$  and insert it into the Lagrangian of the field theory, interpreting the parameter along the path as time.
In practice this means picking a representative (static) field configuration in each equivalence class of solutions of the BPS equations  along the path, thus leading to field configuration $W_{\underline{i}}(t,x^1,x^2,x^3)$ which depend on time and space coordinates.
Time derivatives of such field configurations correspond to tangent vectors along the path,  provided that they satisfy Gauss' Law, which ensures that the time evolution is orthogonal to gauge orbits.
The potential energy on the moduli space is constant, saturating the Bogomol'nyi bound \eqref{BogomolnyiBound}, and using expression \eqref{KV2} for the kinetic energy, the effective Lagrangian
for charge-$\topk$ monopoles is therefore
\begin{align}
L_{\mbox{eff}} & \ = \ \int \d^3 x \ \half \left|\dot{W}_{\underline{i}} \right|^2 - \frac{4\pi a}{e} \Bbk\eqip{,} \label{Lk}
\end{align}
where Gauss' Law  in the form \eqref{GaussLaw} is assumed.

Weinberg\mycite{Weinberg:ParamCount} first argued that the dimension of the moduli space $\MS_{\topk}$ is $4k$, where $k=|k_1+\ldots+k_{n-1}|$ generalising  an  earlier index calculation by Callias\mycite{Callias:78indexThm} (See also Taubes\mycite{Taubes:stabYMT} and Atiyah and Hitchin\mycite{AtiyahHitchin}). Therefore, we can parameterise the moduli space with $4k$ parameters, or moduli, $X^a$. Using these coordinates, tangent vectors to $\MS_{\topk}$ can be decomposed as
\begin{align}
\dot{W}_{\underline{i}} & \ = \ \delta_a W_{\underline{i}} \, \dot{X}^a\eqip{,} \label{tangentvecM}
\end{align}
where the zero-modes $\delta_a W_{\underline{i}}$ form a basis of (lifted) vector fields on $\MS_{\topk}$.

The metric on $\MS_{\topk}$ is obtained by restricting the metric on $\mathcal{A}$. The natural metric on $\mathcal{A}$ is given by
\begin{align}
g(\dot{W}, \dot{V}) & \ = \ \int \d^3 x \ \dot{W}_{\underline{i}} \cdot \dot{V}_{\underline{i}}\eqip{,} \anumber
\end{align}
and the components of its restriction to the moduli space $\MS_{\topk}$, with respect to the basis of zero-modes $\delta_a W_{\underline{i}}$, are
\begin{align}
g_{ab} & \ = \ \int \d^3 x \ \delta_a W_{\underline{i}} \cdot \delta_b W_{\underline{i}}\eqip{.} \label{metricM}
\end{align}
Inserting the decomposition of tangent vectors with respect to the basis of zero-modes \eqref{tangentvecM} into the effective Lagrangian \eqref{Lk}, we have
\begin{align}
L_{\mbox{eff}} & \ = \ \half g(\dot{W}, \dot{W}) - \frac{4\pi a}{e} \Bbk \ = \ \half g_{ab} \dot{X}^a \dot{X}^b - \frac{4\pi a}{e} \Bbk\eqip{.} \label{LeffM}
\end{align}
Because the potential energy on $\MS_{\topk}$ is constant, the equations of motion for this Lagrangian are simply the geodesic equations for the moduli space.
Therefore, slowly moving monopoles follow geodesics on the moduli space\mycite{Manton:ScatteringBPSMon}.

From Gauss' Law, we derive the background gauge condition for the zero-modes,
\begin{align}
 D_{\underline{i}} \delta_a W_{\underline{i}} & \ = \ 0\eqip{,} \label{bgc}
\end{align}
and since field configurations corresponding to points in $\MS_{\topk}$ satisfy the Bogomol'nyi equations, tangent vectors to $\MS_{\topk}$ must satisfy the linearised Bogomol'nyi equations,
\begin{align}
D_{\underline{i}} \dot{W}_{\underline{j}} - D_{\underline{j}} \dot{W}_{\underline{i}} & \ = \ \pm \eps_{\underline{ijkl}} D_{\underline{k}} \dot{W}_{\underline{l}}\eqip{.} \label{linBogomolnyi}
\end{align}

\subsection{The hyperk\"ahler structure of $\MS_{\topk}$}
In the case of maximal symmetry breaking, moduli spaces of monopoles are hyperk\"ahler manifolds. A hyperk\"ahler structure on a Riemannian manifold consists of three parallel complex structures, $\ICS_i$, that obey the quaternion algebra, $\ICS_i \ICS_j = -\delta_{ij} + \eps_{ijk} \ICS_k$. We will sometimes denote the complex structures by
\begin{align}
\ICS & \ = \ \ICS_3\eqip{,} &
\JCS & \ = \ \ICS_1\eqip{,} &
\KCS & \ = \ \ICS_2\eqip{.} \anumber
\end{align}
The existence of a hyperk\"ahler structure on the moduli space is deeply connected to the fact that the field theory allows for an $N=4$ supersymmetric extension\mycite{FigueroaEtAl:SusyHKM}.

The hyperk\"ahler structure on the moduli space derives from the hyperk\"ahler structure on Euclidean $\RR^4$. In the following we use the same symbols $\ICS_i$ for the action of the complex structures on $\RR^4$, on the space of field configurations $\mathcal{A}$, and on the moduli space $\MS_{\topk}$. When we interpret $W_{\underline{i}}$ as a connection on $\RR^4$, the space of field configurations inherits three complex structures from this $\RR^4$, via\mycite{Gauntlett:LowEnDymSusySol, Gauntlett:LowEnDymN=2Mon}
\begin{align}
(\ICS_i (\dot{W}))_{\underline{j}} & \ = \ (\ICS_i)_{\underline{jk}} \ \dot{W}_{\underline{k}}\eqip{.} \label{compStructureAi}
\end{align}
If $\dot{W}_{\underline{j}}$ is a tangent vector to $\MS_{\topk}$, then so is the linear combination $(\ICS_i)_{\underline{jk}} \dot{W}_{\underline{k}}$, and therefore the complex structures on $\mathcal{A}$ can be restricted to complex structures on $\MS_{\topk}$. In terms of the moduli space coordinates, the components of the  complex structures are given by ${(\ICS_i)^a}_b = g^{ac} (\omega_i)_{cb}$, where $\omega_i$ are the K\"ahler forms $\omega_i (\dot{W}, \dot{V}) = \omega_{\ICS_i} (\dot{W}, \dot{V}) = g( \dot{W}, \ICS_i (\dot{V}) )$. The action of the  complex structures on the zero-modes is then found to satisfy
\begin{align}
\delta_b W_{\underline{i}} {(\ICS_i)^b}_a & \ = \ (\ICS_i)_{\underline{ij}} \ \delta_a W_{\underline{j}}\eqip{.} \label{actionCompStructBosZeromode}
\end{align}
The three complex structure again obey the quaternion algebra. Since the K\"ahler forms and the metric are parallel, the complex structures must be as well, and therefore they form a hyperk\"ahler structure on $\MS_{\topk}$. The hyperk\"ahler structure of the moduli space can also be shown by interpreting the moduli space as a hyperk\"ahler quotient\mycite{HitchinEtAl:HKMSuSy,AtiyahHitchin}.

Some of the above statements can be understood most easily  by  combining the bosonic zero-modes into a quaternion as follows
\begin{align}
\qBos_a & \ = \ \delta_a W_4 - \qj \, \delta_a W_1 - \qk \, \delta_a W_2 - \qi \, \delta_a W_3
 \ = \ \qq_{\underline{i}} \, \delta_a W_{\underline{i}}\eqip{,} \label{bzmq}
\end{align}
where $\qq_4 = 1$, and $\qq_1 = \qj$, $\qq_2 = \qk$ and $\qq_3 = \qi$ are the imaginary unit quaternions, satisfying the quaternion algebra $\qi^2 = \qj^2 = \qk^2 = \qi\qj\qk = -1$.

The action of the complex structures \eqref{actionCompStructBosZeromode} can now be expressed as multiplication of the quaternion $\qBos_a$ and its quaternionic conjugate $\overline{\qBos}_a = \delta_a W_4 + \qj \, \delta_a W_1 + \qk \, \delta_a W_2 + \qi \, \delta_a W_3$  with the unit quaternions:
\begin{align}
  \ICS_i(\qBos_a) & \ = \  - \qBos_a\qq_i  \eqip{,} &
  \ICS_i(\overline{\qBos}_a) & \ = \ \qq_i\overline{\qBos}_a \eqip{.} \label{actionCompStructBZMQ}
\end{align}
The components of the metric and the K\"ahler forms on the moduli space are then given by the real and imaginary parts of
\begin{align}
\int \d^3 x \ \qBos_b\overline{\qBos}_a   \ = \ & g_{ab} + \qq_i (\omega_i)_{ab} \label{metricMQ}
\end{align}
respectively, and hence invariant under the action of the complex structures.
The background gauge condition \eqref{bgc} and the linearised Bogomol'nyi equations \eqref{linBogomolnyi} can also be written together in quaternionic form. We define the quaternionic differential operator
\begin{align}
  \qD \ = \ \qq_{\underline{j}} D_{\underline{j}}\eqip{,} \anumber
\end{align}
in terms of which the background gauge condition and the linearised Bogomol'nyi equations become the real and imaginary parts of
\begin{align}
\qD \qBos_a \ = \ 0 \label{linEqnsQ}
\end{align}
respectively. The action of the complex structures \eqref{actionCompStructBZMQ} therefore leaves this set of equations invariant.


\section{N=2 supersymmetric monopoles}
\label{sectN=2}
We now turn to $N=2$ supersymmetric monopoles. As in the bosonic case, we will first review BPS monopoles and the moduli space approximation. Then we will discuss the quantisation of the effective action of the supersymmetric model. This can be done in terms of both spinors (section~ \ref{sectFZMS}), and anti-holomorphic forms on the moduli space (section~\ref{sectFZMF}). We review the realisation of particular supersymmetry charges as a Dirac operator on spinors and a Dolbeault operator on anti-holomorphic forms. As a standard reference for this discussion, we refer to Gauntlett's paper\mycite{Gauntlett:LowEnDymN=2Mon}. A recent review by Weinberg and Yi\mycite{WeinbergYi:MMonDynSusyDuality} discusses these topics in a wider context. The lecture notes on electromagnetic duality by Figueroa-O'Farrill\mycite{Figueroa:edc} are a useful guide for many of the calculations.

To complete this story of quantisation of the effective action, we construct the differential operators corresponding to the remaining supercharges, and interpret them as twisted Dirac operators acting on spinors (in section~\ref{sectFZMSSusy}) and twisted Dolbeault operators acting on anti-holomorphic forms (in section~\ref{sectFZMFSusy}). The identification of all the supercharges is essential for finding all the states in a supermultiplet, as we will illustrate  in sections \ref{sectEx1} and \ref{sectEx2}.

\subsection{N=2 supersymmetric BPS monopoles}
The Yang-Mills-Higgs Lagrangian \eqref{YMHLagrangian} can be extended with $N=2$ supersymmetry. Besides the gauge fields, we now have a scalar field $S$, a pseudo-scalar field $P$ and a fermion field $\psi$. The Lagrangian is given by\mycite{Gauntlett:LowEnDymN=2Mon}
\begin{align}
L 
 & \ = \ \int \d^3 x \ \Big( - \frac{1}{4} F^{\mu\nu} \cdot F_{\mu\nu} + \half D_{\mu} S \cdot D^{\mu} S + \half D_{\mu} P \cdot D^{\mu} P - \half[e^2] || \left[ P, S \right] ||^2 \nonumber \\
 & \qquad \qquad \qquad + i \overline{\psi} \cdot \gamma^{\mu} D_{\mu} \psi + i e \overline{\psi} \cdot \left[ S -i \gamma_5 P, \psi \right] \Big)\eqip{,} \label{LN=2}
\end{align}
where $\gamma_5 = i \gamma_0 \gamma_1 \gamma_2 \gamma_3$. We take the $\gamma$-matrices to satisfy $\{ \gamma_{\mu}, \gamma_{\nu} \} = 2 \eta_{\mu\nu}$.
This Lagrangian can be derived from an $N=1$ supersymmetric model in six dimensions, via dimensional reduction\mycite{BrinkSchwSch:SusyYMT,DAddaEtAl:SuSyMMD}. The rotational symmetry of the extra dimensions reduces to a chiral rotational symmetry in four dimensions. As in the bosonic case, the symmetry breaking is induced by choosing appropriate boundary conditions on the fields at infinity: $\lim_{r\to\infty} ||S||^2 + ||P||^2 = a^2$.
%

To find the zero-modes we first use a chiral rotation, so that we may assume that only the scalar field $S$ has a non-zero vacuum expectation value. In this case $S$ takes on the role of the Higgs field of the bosonic model.
The kinetic and potential energy for the supersymmetric model are
\begin{align}
K & \ = \ \int \d^3 x \ \Big( - || \half F_{0i} ||^2 + \half || D_{0} S ||^2 + \half || D_{0} P ||^2 + i \overline{\psi} \cdot \gamma^{0} D_{0} \psi \Big)\eqip{,} \anumber \\
V & \ = \ \int \d^3 x \ \Big( \ \frac{1}{4} F^{ij} \cdot F_{ij} + \half || D_{i} S ||^2 + \half || D_{i} P ||^2 + \half[e^2] || \ \left[ P, S \right] \ ||^2 \nonumber \\
 & \qquad \qquad \qquad + i \overline{\psi} \cdot \gamma_{i} D_{i} \psi - i e \overline{\psi} \cdot \left[ S -i \gamma_5 P, \psi \right] \Big)\eqip{.} \anumber
\end{align}
The BPS monopoles are defined, as before, to have minimal energy. This implies again that they are static. Furthermore, assuming $S$ to be the only field with a non-zero vacuum expectation value, $S$ must satisfy the Bogomol'nyi equations \eqref{Bogomolnyi}, and $\psi$ must satisfy the following Dirac equation in the presence of a monopole background,
\begin{align}
\gamma_0 \gamma_{i} D_{i} \psi - e \gamma_0 \left[ S, \psi \right] & \ = \ 0\eqip{.} \label{diracFZMa}
\end{align}


We define Euclidian gamma-matrices by
\begin{align}
\Egamma_i & \ = \ \gamma_0 \gamma_i\eqip{,} &
\Egamma_4 & \ = \ \gamma_0\eqip{,} \label{defgammabar}
\end{align}
which satisfy $\{ \Egamma_{\underline{i}} , \Egamma_{\underline{j}} \} = 2 \delta_{\underline{ij}}$. In terms of these, the Dirac equation \eqref{diracFZMa} becomes
\begin{align}
\slD \psi \ \equiv \ \Egamma_{\underline{i}} D_{\underline{i}} \psi & \ = \ 0\eqip{,} \label{diracFZM}
\end{align}
where $D_{\underline{i}}$ is the covariant derivative in Euclidian space defined in equation \eqref{defD}.

\subsection{Zero-modes}
\label{sectFZeroModes}
The bosonic zero-modes of this model are exactly the same as those of the purely bosonic model. The fermionic zero-modes are the solutions of the Dirac equation \eqref{diracFZM}. A convenient representation for the Euclidian $\gamma$-matrices is given by
\begin{align}
  \Egamma_i & \ = \ \left(
                                \begin{array}{cc}
                                  0 & i \sigma_i \\
                                  -i \sigma_i & 0 \\
                                \end{array}
                              \right)\eqip{,} &
  \Egamma_4 & \ = \ \left(
                                \begin{array}{cc}
                                  0 & \unit_2 \\
                                  \unit_2 & 0 \\
                                \end{array}
                              \right)\eqip{.} \label{euclidGammaMatr}
\end{align}
Using this representation, and identifying the imaginary unit quaternions with the Pauli matrices,
\begin{align}
  \qq_1 
    & \ \sim \ -i\sigma_1\eqip{,} &
  \qq_2 
    & \ \sim \ -i\sigma_2\eqip{,} &
  \qq_3 
    & \ \sim \ -i\sigma_3\eqip{,} &
  \qq_4 
    & \ \sim \ \unit_2\eqip{,} \anumber
\end{align}
the Dirac equation \eqref{diracFZM} becomes
\begin{align}
  \left(
  \begin{array}{cc}
    0 & \qD^{\dag} \\
    \qD & 0 \\
  \end{array}
\right) \psi \ = \ 0\eqip{,} \anumber
\end{align}
where $\qD$ is defined in \eqref{linEqnsQ}. Since $\qD^{\dag}$ has no normalisable zero-modes, the fermionic zero-modes can all be written in terms of the bosonic zero-modes as\mycite{Gauntlett:LowEnDymN=2Mon}
\begin{align}
\psi_a & \ = \ \Egamma_{\underline{i}} \delta_a W_{\underline{i}} \left(
    \begin{array}{c}
      0 \\
      \chi \\
    \end{array}
  \right)
     \ = \ \left(
             \begin{array}{c}
               \overline{\qq}_{\underline{i}} \delta_a W_{\underline{i}} \ \chi \\
               0 \\
             \end{array}
           \right)
          \ = \ \left(
             \begin{array}{c}
               \qBos_a \, \chi \\
               0 \\
             \end{array}
           \right)\eqip{,} \label{fermZM}
\end{align}
where the bosonic zero-mode $\overline{\qBos}_a$ was defined by equation \eqref{bzmq}, and $\chi$ is a fixed, normalised two-component spinor.

The complex structures act on the fermionic zero-modes \eqref{fermZM} via equation \eqref{actionCompStructBosZeromode} or \eqref{actionCompStructBZMQ}.
Using hats to distinguish the action of the complex structures on fermionic zero-modes from the action of the complex structures on bosonic zero-modes, we have
\begin{align}
\hat{\ICS}_i(\psi_a) & \ = \ \left(
                      \begin{array}{c}
                        \ICS_i(\qBos_a) \chi \\
                        0 \\
                      \end{array}
                    \right) \ = \ \left(
                      \begin{array}{c}
                        -\qBos_a \qq_i \chi \\
                        0 \\
                      \end{array}
                    \right)\eqip{.} \label{actionCompStructFermZeromode}
\end{align}
We see that there is a complex structure $\hat{\ICS}$ (which depends on the choice of $\chi$) such that\mycite{Gauntlett:LowEnDymN=2Mon}
\begin{align}
\hat{\ICS} (\psi_a) & \ = \ i \psi_a\eqip{.} \label{IFZM}
\end{align}
For example, if we choose $\chi = \left(
    \begin{array}{c}
      1 \\
      0 \\
    \end{array}
  \right)$, then $\hat{\ICS} = \hat{\ICS}_3$.
Having made a choice, and fixed $\chi$, the remaining two complex structures that make up the hyperk\"ahler structure, $\hat{\JCS}$ and $\hat{\KCS}$ act anti-linearly:
\begin{align}
\hat{\JCS} ( i \psi_a ) & \ = \ \hat{\JCS} \hat{\ICS} (\psi_a) \ = \ - \hat{\ICS} \hat{\JCS} (\psi_a) \ = \ - i \hat{\JCS} (\psi_a)\eqip{,} \anumber
\end{align}
and similarly for $\hat{\KCS}$.

Equation \eqref{IFZM} shows that the $4k$ fermionic zero-modes $\psi_a$, defined by \eqref{fermZM} in terms of the $4k$ bosonic zero-modes, are not linearly independent over $\CC$. The $4k$ $\RR$-dimensional vector space of bosonic zero-modes corresponds to a $2k$ $\CC$-dimensional vector space $V$ of fermionic zero-modes, in agreement with Callias' index theorem\mycite{Callias:78indexThm}.

$V$ can also be viewed as a $4k$ $\RR$-dimensional vector space. If
\begin{align}
\mathscr{B}_V^{\CC} \ = \ \{\psi_1, \ldots, \psi_{2k}\} \label{CbasisV}
\end{align}
is a basis of $V$ over $\CC$, then, in light of equation \eqref{IFZM}, a basis of $V$ over $\RR$ is given by
\begin{align}
\mathscr{B}_V^{\RR} \ = \ \{\psi_1, \ldots, \psi_{2k}, \hat\ICS(\psi_1),  \ldots,\hat \ICS(\psi_{2k})\}\eqip{.} \label{RbasisV}
\end{align}


\subsection{Fermionic zero-modes as anti-holomorphic forms on $\RR^4$}
\label{sectFZMFR4}
The fermionic zero-modes $\psi_a$ \eqref{fermZM} are static spinors in space-time. Extending space-time to $\RR \times \RR^4$, as we did in section~\ref{E4}, we may also view them as spinors on $\RR^4$ that are independent of the fourth dimension. These spinors can now be identified with anti-holomorphic forms\mycite{Hitchin:HarmSpin}. For example, we can identify the fermionic zero-modes $\psi_a$ with anti-holomorphic forms $\overline{\upsilon}_a$ via
\begin{align}
\label{spinformid}
  \psi_a \ = \ \left(
                   \begin{array}{c}
                     a_1 \\
                     a_2 \\
                     0 \\
                     0 \\
                   \end{array}
                 \right)
  \qquad \sim \qquad &
  \overline{\upsilon}_a \ = \ a_1 \overline{\alpha}_1 + a_2 \overline{\alpha}_2\eqip{,} \anumber
\end{align}
where  we use the basis $\overline{\alpha}_1=\frac{1}{\sqrt{2}}(\d x^3-i \d x^4)$ and $\overline{\alpha}_2=\frac {1}{\sqrt{2}}(\d x^1-i \d x^2)$  of anti-holomorphic forms (with respect to the complex structure $\ICS$) on $\RR^4$. The complex structures $\ICS,\JCS,\KCS$ act naturally on the space of all differential forms, and hence on this basis. One finds
\begin{align}
  \ICS(\overline{\alpha}_1) & \ = \ i \overline{\alpha}_1\eqip{,} &
  \JCS(\overline{\alpha}_1) & \ = \ \phantom{-} i \alpha_2\eqip{,} &
  \KCS(\overline{\alpha}_1) & \ = \ \phantom{-} \alpha_2\eqip{,} \nonumber \\
  \ICS(\overline{\alpha}_2) & \ = \ i \overline{\alpha}_2\eqip{,} &
  \JCS(\overline{\alpha}_2) & \ = \ -i \alpha_1\eqip{,} &
  \KCS(\overline{\alpha}_2) & \ = \ - \alpha_1\eqip{.} \anumber
\end{align}
We would like to clarify how these (linear) actions of the complex structures on  forms are related to the (linear) action of $\hat{\ICS}$, and the (anti-linear) actions of $\hat{\JCS}$ and $\hat{\KCS}$ on spinors defined in \eqref{actionCompStructFermZeromode}.  Using the identification  \eqref{spinformid}, we can pull the maps $\hat{\ICS}$, $\hat{\JCS}$ and $\hat{\KCS}$ back to maps on anti-holomorphic forms. Denoting these pull-backs by the same letters $\hat{\ICS}$, $\hat{\JCS}$ and $\hat{\KCS}$ we find
\begin{align}
  \hat{\ICS}(\overline{\upsilon}_a) & \ = \ \phantom{i \, } \ICS(\overline{\upsilon}_a) \ = \ i \, \overline{\upsilon}_a\eqip{,} &
  \hat{\ICS} & \ = \ \phantom{i \, } \ICS\eqip{,} \nonumber \\
  \hat{\JCS}(\overline{\upsilon}_a) & \ = \ - \, \overline{\JCS}(\overline{\upsilon}_a) \ = \ - \, \overline{\JCS(\overline{\upsilon}_a)}\eqip{,} &
  \hat{\JCS} & \ = \ - \, \overline{\JCS}\eqip{,} \nonumber \\
  \hat{\KCS}(\overline{\upsilon}_a) & \ = \ - \, \overline{\KCS}(\overline{\upsilon}_a) \ = \ - \, \overline{\KCS(\overline{\upsilon}_a)}\eqip{,} &
  \hat{\KCS} & \ = \ - \, \overline{\KCS}\eqip{,} \label{actionCSFZF}
\end{align}
where $\overline{\JCS}$ and $\overline{\KCS}$ are anti-linear maps from the space of anti-holomorphic forms to itself. Of course,  $\hat{\ICS}$, $\hat{\JCS}$ and $\hat{\KCS}$ still obey the quaternion algebra.

\subsection{Effective Lagrangian}
Viewing the space of fermionic zero-modes as a $4k$ $\RR$-dimensional vector space, we parameterise the fermionic zero-modes as
\begin{align}
\psi & \ = \ \psi_a \lambda^a\eqip{,} &
\psi_a & \in \mathscr{B}_V^{\RR}\eqip{,} &
\lambda^a & \in \RR\eqip{.} \label{FZMparam}
\end{align}
Since $\psi_a$ is defined to be a commuting spinor, $\lambda^a$ is a Grassmann variable.

Inserting this parametrisation into the Lagrangian \eqref{LN=2} and expanding to lowest non-trivial order, Gauntlett\mycite{Gauntlett:LowEnDymN=2Mon} found
\begin{align}
L_{\mbox{eff}} & \ = \ \half \left[ g(\dot{X}, \dot{X}) + i g(\lambda, D_t \lambda) \right] - \frac{4\pi a}{e} \Bbk\eqip{,} \label{LeffN=2}
\end{align}
where the covariant derivative $D_t = \dot{X}^a D_a$, and
\begin{align}
(D_t \lambda)^a & \ = \ \dot{\lambda}^a + \Gamma^a_{bc} \dot{X}^b \lambda^c\eqip{.} \anumber
\end{align}

On our way to a quantum mechanical description of the $N=2$ supersymmetric monopoles at low energies we may now proceed in two different, but equivalent ways. If we continue to work with real coordinates $X^a$ on the moduli space, we naturally end up with a quantum theory of spinors on the moduli space (section~\ref{sectFZMS}). Alternatively, if we choose to work with complex coordinates, the natural way to quantise the theory leads to anti-holomorphic forms on the moduli space (section~ \ref{sectFZMF}). Since the moduli space is hyperk\"ahler, and hence Ricci flat, these two descriptions are equivalent\mycite{Gauntlett:LowEnDymSusySol}, as we shall demonstrate explicitly for the examples $\MS_1$ and $\MS_{1,1}$ in sections \ref{sectEx1} and \ref{sectEx2}.

\section{Fermionic zero-modes as spinors on the moduli space}
\label{sectFZMS}
\subsection{Effective Hamiltonian, using real coordinates}
The quickest route towards quantisation is to start by introducing an orthonormal frame to parameterise the fermionic zero-modes (Friedan and Windey\mycite{FrieWin:SusyDerAtiSinInd}, Davis, Macfarlane, Popat and Van Holten\mycite{DavisEtAl:qmSusyNLSigma, MacfPop:qmN=2SusyNLSM}, and Gauntlett\mycite{Gauntlett:LowEnDymSusySol}). The effective Hamiltonian can be derived without such a frame, but the canonical momenta one finds in this case are not suitable for quantisation, and an orthonormal frame will have to be introduced eventually. We define the orthonormal frame $\ofr$ by
\begin{align}
g_{ab} & \ = \ \delta_{AB} {\ofr^A}_a {\ofr^B}_b\eqip{,} &
\ofr^A & \ = \ {\ofr^A}_a \d X^a\eqip{,} &
\lambda^A & \ = \ {\ofr^A}_a \lambda^a\eqip{.} \anumber
\end{align}
In terms of the orthonormal frame, the covariant derivative of $\lambda$ becomes
\begin{align}
(D_t \lambda)^A & \ = \ \dot{\lambda}^A + {{\omega_a}^A}_B \dot{X}^a \lambda^B\eqip{,} \anumber
\end{align}
where the spin connection $\omega$ is determined by a gauge transformation
\begin{align}
{{\omega_a}^A}_B & \ = \ {\ofr^A}_b \Gamma^b_{ac} {\ofr^c}_B + {\ofr^A}_b \partial_a {\ofr^b}_B\eqip{,} \anumber
\end{align}
where ${\ofr^c}_B$ is the inverse of ${\ofr^B}_c$, in the sense that ${\ofr^c}_B {\ofr^B}_d = \delta^c_d$ and ${\ofr^A}_c {\ofr^c}_B = \delta^A_B$.

The effective Lagrangian \eqref{LeffN=2} can therefore be written as
\begin{align}
L_{\mbox{eff}} & \ = \ \half \left[ g_{ab} \dot{X}^a \dot{X}^b + i \delta_{AB} \lambda^A (D_t \lambda)^B \right] - \frac{4\pi a}{e} \Bbk\eqip{.} \anumber
\end{align}
The canonical momenta are then
\begin{align}
p_a \ = \ \frac{\partial L_{\mbox{eff}}}{\partial \dot{X}^a} & \ = \ g_{ab} \dot{X}^b + \half[i] {\omega_a}_{AB} \lambda^A \lambda^B\eqip{,} &
\pi_A \ = \ \frac{\partial L_{\mbox{eff}}}{\partial \dot{\lambda}^A} & \ = \ - \half[i] \delta_{AB} \lambda^B\eqip{,} \anumber
\end{align}
where ${\omega_a}_{AB} = \delta_{AC} {{\omega_a}^C}_B$. The effective Hamiltonian is given by
\begin{align}
H_{\mbox{eff}} & \ = \ \dot{X}^a p_a + \dot{\lambda}^A \pi_A - L_{\mbox{eff}} \ = \ H_0 + \frac{4\pi a}{e} \Bbk\eqip{,} \anumber
\end{align}
where we have defined
\begin{align}
H_0 & \ = \ \half g^{ab} \tilde{p}_a \tilde{p}_b\eqip{,} &
\tilde{p}_a & \ = \ p_a - \half[i] {\omega_a}_{AB} \lambda^A \lambda^B \ = \ g_{ab} \dot{X}^b\eqip{.} \anumber
\end{align}

\subsection{Quantisation}
Generally,  a symmetry is generated by its corresponding charge through a Poisson bracket. In this case, however, the expression for the fermionic momenta leads to constraints. We must therefore replace Poisson brackets by Dirac brackets. Having used an orthonormal frame to define the fermionic variables $\lambda^A$, we find that the Dirac brackets of the bosonic and fermionic variables decouple. The only non-vanishing Dirac brackets are
\begin{align}
  \left\{ p_a , X^b \right\}_{DB} & \ = \ \delta_a^b\eqip{,} &
  \left\{ \lambda^A , \lambda^B \right\}_{DB} & \ = \ i \delta^{AB}\eqip{.} \anumber
\end{align}
To quantise the theory we follow Friedan and Windey\mycite{FrieWin:SusyDerAtiSinInd}, Alvarez-Gaum\'e\mycite{AlvarezGaume:SusyAtiSinIndTh}, and Gauntlett\mycite{Gauntlett:LowEnDymN=2Mon}. Dirac brackets of bosons are replaced with commutators, while Dirac brackets of fermions are replaced with anti-commutators.
\begin{align}
\left\{ p_a , X^b \right\}_{DB} & \ = \ \phantom{i} \delta_a^b & \mapsto &&
\left[ \hat{p}_a , \hat{X}^b \right] & \ = \ -i \delta_a^b \anumber \\
\left\{ \lambda^A , \lambda^B \right\}_{DB} & \ = \ i \delta^{AB} & \mapsto &&
\left\{ \hat{\lambda}^A , \hat{\lambda}^B \right\} & \ = \ \phantom{-i} \delta^{AB} \anumber
\end{align}
The anti-commutator of the fermions defines a Clifford bundle over $\MS_{\topk}$, and we identify the Hilbert space of states with the space of spinors on $\MS_{\topk}$. The bosonic coordinates act by multiplication and the bosonic momenta are represented as derivatives,
\begin{align}
p_a & \ \mapsto \ -i \partial_a\eqip{.} \anumber
\end{align}
We have a natural map from the fermions to the Dirac matrices on the moduli space:
\begin{align}
\lambda^A  \ \mapsto \ & \frac{i}{\sqrt{2}} \gamma^A\eqip{,} &
\{ \gamma^A, \gamma^B \} & \ = \ -2 \delta^{AB}\eqip{.} \label{lambdaQgamma}
\end{align}
We see that $\tilde{p}_a$ acts as the covariant derivative on spinors,
\begin{align}
\tilde{p}_a \ = \ & p_a - \frac{i}{4} {\omega_a}_{AB} \left[ \lambda^A, \lambda^B \right] & \mapsto && -i \left( \partial_a - \frac{1}{8} {\omega_a}_{AB} \left[ \gamma^A , \gamma^B \right] \right) \ = \ -i D_a\eqip{.} \anumber
\end{align}
Finally, the quantisation of the effective Hamiltonian $H_0$ gives half the Laplacian,
\begin{align}
H_0 & \ = \ \half g^{ab} \tilde{p}_a \anumber \tilde{p}_b & \mapsto && - \half g^{ab} D_a D_b \ = \ \half \Delta\eqip{.} \anumber
\end{align}
Here we have defined the Laplacian to be the positive definite operator, which corresponds to the usual definition for the Laplacian acting on forms, $\Delta = (\d + \d^{\dag})^2$.

\subsection{Supersymmetry}
\label{sectFZMSSusy}
The effective action \eqref{LeffN=2} is invariant under $N=4$ supersymmetry transformations\mycite{Gauntlett:LowEnDymN=2Mon}:
\begin{align}
\delta_{\unit} X^a & \ = \ \phantom{i}\eps \lambda^a &
\delta_{\ICS_j} X^a & \ = \ \eps \ {(\ICS_j)^a}_b \lambda^b \nonumber \\
\delta_{\unit} \lambda^a & \ = \ i \eps \dot{X}^a 
&
\delta_{\ICS_j} \lambda^a & \ = \ \eps \left[ - i {(\ICS_j)^a}_b \dot{X}^b - \Gamma^a_{cd} {(\ICS_j)^c}_b \lambda^b \lambda^d \right] \anumber
\end{align}
which have their origin in the unbroken supersymmetries of the field theory.
The corresponding supercharges are
\begin{align}
Q_{\unit} & \ = \ \tilde{p}_a \lambda^a\eqip{,} &
Q_{\ICS_i} & \ = \ \tilde{p}_b {(\ICS_i)^b}_a \lambda^a\eqip{.} \anumber
\end{align}
The supercharges generate the supersymmetry transformations via Dirac brackets, and they obey the $N=4$ supersymmetry algebra:
\begin{align}
\left\{ Q_{\unit}, Q_{\unit} \right\}_{DB} & \ = \ 2 i H_0\eqip{,} &
\left\{ Q_{\ICS_i}, Q_{\ICS_j} \right\}_{DB} & \ = \ \delta_{ij} \ 2 i H_0\eqip{,} \anumber
\end{align}
with all other brackets vanishing. This agrees with the fact that the supersymmetry transformations square to $i$ times a time-derivative, and that time evolution is generated by the Hamiltonian.

When we quantise the supercharges using the quantisation procedures given above, $Q_{\unit}$ becomes the Dirac operator for spinors on the moduli space
\begin{align}
Q_{\unit} & \ = \ \tilde{p}_a \lambda^a & \mapsto &&
  \frac{1}{\sqrt{2}} \gamma^a D_a & \ = \ \frac{1}{\sqrt{2}} \, \slD\eqip{,} \anumber
\intertext{while the remaining supercharges become twisted Dirac operators}
Q_{\ICS_j} & \ = \ \tilde{p}_b {(\ICS_j)^b}_a \lambda^a & \mapsto &&
  \frac{1}{\sqrt{2}} {(\ICS_j)^b}_a \gamma^a D_b & \ =: \ \frac{1}{\sqrt{2}} \, \slD_{\ICS_j}\eqip{.} \anumber
\end{align}
From the fact that the Hamiltonian is given by the Dirac bracket of supercharges, we find again that the quantisation of $H_0$ gives half the Laplacian:
\begin{align}
  H_0 & \ = \ -\half[i] \left\{ Q_{\unit}, Q_{\unit} \right\}_{DB} & \mapsto &&
  \half \slD^2 & \ = \ - \half g^{ab} D_a D_b \ = \ \half \Delta\eqip{.} \anumber
\end{align}

\section{Fermionic zero-modes as forms on the moduli space}
\label{sectFZMF}
\subsection{Effective Hamiltonian, using complex coordinates}
We now take a few steps back to discuss the quantisation of the effective model on the moduli space in terms of anti-holomorphic forms on the moduli space. We choose $2k$ (local) complex coordinates $Z^{\alpha}$ for the hyperk\"ahler manifold $\MS_{\topk}$ with respect to the complex structure $\ICS = \ICS_3$. The real and imaginary parts of $Z^{\alpha}$ form a basis of real coordinates $X^a$ (the index $a$, as usual, runs from $1$ to $4k$).
We may choose these coordinates  so  that
\begin{align}
  Z^{\alpha} \ = \ & X^{\alpha} + i X^{\alpha + 2k}\eqip{,} &
  \overline{Z}^{\overline{\alpha}} \ = \ & X^{\alpha} - i X^{\alpha + 2k}\eqip{.} \anumber
\end{align}
Equation \eqref{IFZM} implies that in the basis $\mathscr{B}_V^{\RR}$ \eqref{RbasisV} we have $\psi_{\alpha + 2k} = \ICS(\psi_{\alpha}) = i \psi_{\alpha}$. Therefore, when we view the space of fermionic zero-modes as a $2k$ $\CC$-dimensional vector space with basis $\mathscr{B}_V^{\CC}$ \eqref{CbasisV}, the fermionic zero-modes are parameterised by
\begin{align}
  \psi \ = \ & \psi_a \lambda^a & a \in & \left\{1, \ldots, 4k\right\} \nonumber \\
       \ = \ & \psi_{\alpha} \left(\lambda^{\alpha} + i \lambda^{\alpha+2k}\right) & \alpha \in & \left\{1, \ldots, 2k\right\}\eqip{.} \anumber
\end{align}
We now define
\begin{align}
  \zeta^{\alpha} \ = \ & ( \lambda^{\alpha} + i \lambda^{\alpha + 2k} )\eqip{,} &
  \zeta^{\overline{\alpha}} \ = \ &  ( \lambda^{\alpha} - i \lambda^{\alpha + 2k} )\eqip{.} \anumber
\end{align}
The fermionic zero-modes are then parameterised by $\psi =  \psi_{\alpha} \zeta^{\alpha}$. The effective Lagrangian in terms of these coordinates becomes
\begin{align}
L_{\mbox{eff}} & \ = \ g_{\alpha \overline{\beta}} \dot{Z}^{\alpha} \dot{\overline{Z}}^{\overline{\beta}} + i g_{\overline{\alpha} \beta} \zeta^{\overline{\alpha}} \left( \dot{\zeta}^{\beta} + \Gamma^{\beta}_{\gamma\delta} \dot{Z}^{\gamma} \zeta^{\delta} \right) - \frac{4\pi a}{e} \Bbk\eqip{.} \anumber
\end{align}

We introduce an orthonormal frame $\ofc$ for the fermionic variables, choosing it so that it respects holomorphicity\mycite{Gauntlett:LowEnDymSusySol}.
\begin{align}
g_{\alpha\overline{\beta}} & \ = \ \delta_{A\overline{B}} {\ofc^A}_{\alpha} {\ofc^{\overline{B}}}_{\overline{\beta}} &
\ofc^A & \ = \ {\ofc^A}_{\alpha} \d Z^{\alpha} &
\zeta^A & \ = \ {\ofc^A}_{\alpha} \zeta^{\alpha} \nonumber \\
&& \ofc^{\overline{A}} & \ = \ {\ofc^{\overline{A}}}_{\overline{\alpha}} \d \overline{Z}^{\overline{\alpha}} &
\zeta^{\overline{A}} & \ = \ {\ofc^{\overline{A}}}_{\overline{\alpha}} \zeta^{\overline{\alpha}} \anumber
\end{align}
The effective Lagrangian then becomes
\begin{align}
L_{\mbox{eff}} & \ = \ g_{\alpha \overline{\beta}} \dot{Z}^{\alpha} \dot{\overline{Z}}^{\overline{\beta}} + i \delta_{\overline{A} B} \zeta^{\overline{A}} \left( \dot{\zeta}^{B} + {{\omega_{\overline{\alpha}}}^{B}}_{C} \dot{\overline{Z}}^{\overline{\alpha}} \zeta^{C} + {{\omega_{\alpha}}^{B}}_{C} \dot{Z}^{\alpha} \zeta^{C} \right) - \frac{4\pi a}{e}\Bbk\eqip{,} \anumber
\end{align}
where the spin connection $\omega$ is again determined by a gauge transformation,
\begin{align}
{{\omega_{\alpha}}^A}_B & \ = \ {\ofc^A}_{\beta} \Gamma^{\beta}_{\alpha\gamma} {\ofc^{\gamma}}_B + {\ofc^A}_{\beta} \partial_{\alpha} {\ofc^{\beta}}_B\eqip{,} &
{{\omega_{\overline{\alpha}}}^A}_B & \ = \ {\ofc^A}_{\beta} \partial_{\overline{\alpha}} {\ofc^{\beta}}_B\eqip{.} \anumber
\end{align}

We compute the canonical momenta from the Lagrangian.
\begin{align}
P_{\alpha} \ = \ \frac{\partial L_{\mbox{eff}}}{\partial \dot{Z}^{\alpha}} & \ = \ g_{\alpha \overline{\beta}} \dot{\overline{Z}}^{\overline{\beta}} + i \omega_{\alpha\overline{A}C} \zeta^{\overline{A}} \zeta^C &
\Pi_{A} \ = \ \frac{\partial L_{\mbox{eff}}}{\partial \dot{\zeta}^{A}} & \ = \ - i \delta_{\overline{B}A} \zeta^{\overline{B}} \nonumber \\
P_{\overline{\alpha}} \ = \ \frac{\partial L_{\mbox{eff}}}{\partial \dot{\overline{Z}}^{\overline{\alpha}}} & \ = \ g_{\beta \overline{\alpha}} \dot{Z}^{\beta} + i \omega_{\overline{\alpha}\overline{A}C} \zeta^{\overline{A}} \zeta^{C} &
\Pi_{\overline{A}} \ = \ \frac{\partial L_{\mbox{eff}}}{\partial \dot{\zeta}^{\overline{A}}} & \ = \ 0 \anumber
\end{align}
The effective Hamiltonian is
\begin{align}
H_{\mbox{eff}} & \ = \ \dot{Z}^{\alpha} P_{\alpha} + \dot{Z}^{\overline{\alpha}} \overline{P}_{\overline{\alpha}} + \dot{\zeta}^{A} \Pi_{A} - L_{\mbox{eff}} \ = \ H_0 + \frac{4\pi a}{e} \Bbk\eqip{,} \anumber
\end{align}
where we have defined
\begin{align}
H_0 & \ = \ g^{\alpha \overline{\beta}} \tilde{P}_{\alpha} \tilde{P}_{\overline{\beta}}\eqip{,} &
\tilde{P}_{\alpha} & \ = \ P_{\alpha} - i \omega_{\alpha\overline{A}C} \zeta^{\overline{A}} \zeta^C\eqip{,} &
\tilde{P}_{\overline{\alpha}} & \ = \ P_{\overline{\alpha}} - i \omega_{\overline{\alpha}\overline{A}C} \zeta^{\overline{A}} \zeta^{C}\eqip{.} \anumber
\end{align}

\subsection{Quantisation}
Just like in section~\ref{sectFZMS}, where we used real coordinates, we must use Dirac brackets in order to quantise the theory, because of the constraints that arise from the expressions for the fermionic momenta. As before, having used an orthonormal frame for the fermionic variables, we find that the brackets of the bosonic and fermionic variables decouple. The only non-vanishing Dirac brackets are
\begin{align}
\left\{ P_{\alpha} , Z^{\beta} \right\}_{DB} & \ = \ \delta_{\alpha}^{\beta} &
\left\{ P_{\overline{\alpha}} , \overline{Z}^{\overline{\beta}} \right\}_{DB} & \ = \ \delta_{\overline{\alpha}}^{\overline{\beta}} &
\left\{ \zeta^A , \zeta^{\overline{B}} \right\}_{DB} & \ = \ i \delta^{A\overline{B}} \anumber
\end{align}
To quantise the theory we again follow the usual procedure. Dirac brackets of bosons are replaced with commutators, and Dirac brackets of fermions are replaced with anti-commutators.
\begin{align}
\left\{ P_{\alpha} , Z^{\beta} \right\}_{DB} & \ = \ \delta_{\alpha}^{\beta} & \mapsto && \left[ P_{\alpha} , Z^{\beta} \right] & \ = \ -i \delta_{\alpha}^{\beta} \nonumber \\
\left\{ P_{\overline{\alpha}} , \overline{Z}^{\overline{\beta}} \right\}_{DB} & \ = \ \delta_{\overline{\alpha}}^{\overline{\beta}} & \mapsto && \left[ P_{\overline{\alpha}} , \overline{Z}^{\overline{\beta}} \right] & \ = \ - i \delta_{\overline{\alpha}}^{\overline{\beta}} \nonumber \\
\left\{ \zeta^A , \zeta^{\overline{B}} \right\}_{DB} & \ = \ i \delta^{A\overline{B}} & \mapsto && \left\{ \zeta^A , \zeta^{\overline{B}} \right\} & \ = \ \delta^{A\overline{B}} \anumber
\end{align}
We can interpret the Hilbert space of states as the space of square-integrable $(0,p)$-forms on $\MS$.
The bosonic coordinates act by multiplication and the bosonic momenta are represented as derivatives,
\begin{align}
P_{\alpha} & \ \mapsto \ -i \partial_{\alpha} &
P_{\overline{\alpha}} & \ \mapsto \ - i \partial_{\overline{\alpha}} \anumber
\end{align}
while while the quantisation of fermions is given by
\begin{align}
\zeta^{\overline{A}} & \ \mapsto \ \ofc^{\overline{A}} \wedge &
\zeta^{A} & \ \mapsto \ \iota(\ofc^{A}), \anumber
\end{align}
where $\iota(\ofc^{A})(\ofc^{\overline{B}}) = \delta^{A \overline{B}}$.
The $\tilde{P}$ act as covariant derivatives,
\begin{align}
\tilde{P}_{\alpha} & \ = \ P_{\alpha} - i \omega_{\alpha\overline{A}C} \zeta^{\overline{A}} \zeta^C & \mapsto &&
-i \left( \partial_{\alpha} + \omega_{\alpha\overline{A}C} \ofc^{\overline{A}} \wedge \iota(\ofc^C) \right) & \ = \ -i \nabla_{\alpha}\eqip{,} \anumber \\
\tilde{P}_{\overline{\alpha}} & \ = \ P_{\overline{\alpha}} - i \omega_{\overline{\alpha}\overline{A}C} \zeta^{\overline{A}} \zeta^{C}
 & \mapsto &&
- i \left( \partial_{\overline{\alpha}} + \omega_{\overline{\alpha}\overline{A}C} \ofc^{\overline{A}} \wedge \iota(\ofc^{C}) \right) & \ = \ - i \nabla_{\overline{\alpha}}\eqip{,} \anumber
\end{align}
and the quantisation of the effective Hamiltonian gives again half the Laplacian,
\begin{align}
H_0 & \ = \ g^{\alpha \overline{\beta}} \tilde{P}_{\alpha} \tilde{P}_{\overline{\beta}} & \mapsto &&
 - g^{\alpha \overline{\beta}} \nabla_{\alpha} \nabla_{\overline{\beta}} \ = \ \half \Delta\eqip{.} \anumber
\end{align}
The upshot of the quantisation is that  the  spinorial zero-modes of the original field theory are represented by anti-holomorphic forms on the moduli space. The quantisation also automatically allows for the possibility that  multiple fermionic zero-modes are excited. Such excitations  are represented by wedge products of anti-holomorphic forms, with the antisymmetric nature of the wedge product reflecting the fermionic nature of the spinor zero-modes.

\subsection{Supersymmetry}
\label{sectFZMFSusy}
The effective action is invariant under $N=4$ supersymmetry transformations\mycite{Figueroa:edc}
\begin{align}
\delta_{\unit} Z^{\alpha} & \ = \ \phantom{-i} \tilde{\eps} \zeta^{\alpha} &
\delta_{\unit} \zeta^{\alpha} & \ = \ \phantom{-} i \tilde{\eps} \dot{Z}^{\alpha} \nonumber \\
\delta_{\unit} \overline{Z}^{\overline{\alpha}} & \ = \ \phantom{-i} \tilde{\eps} \zeta^{\overline{\alpha}} &
\delta_{\unit} \zeta^{\overline{\alpha}} & \ = \ \phantom{-} i \tilde{\eps} \dot{\overline{Z}}^{\overline{\alpha}} \anumber \\
\delta_\ICS Z^{\alpha} & \ = \ \phantom{-}i \tilde{\eps} \zeta^{\alpha} &
\delta_\ICS \zeta^{\alpha} & \ = \ \phantom{-i} \tilde{\eps} \dot{Z}^{\alpha} \nonumber \\
\delta_\ICS \overline{Z}^{\overline{\alpha}} & \ = \ -i \tilde{\eps} \zeta^{\overline{\alpha}} &
\delta_\ICS \zeta^{\overline{\alpha}} & \ = \ -\phantom{i} \tilde{\eps} \dot{\overline{Z}}^{\overline{\alpha}} \anumber \\
\delta_\JCS Z^{\alpha} & \ = \ \tilde{\eps} {\JCS^{\alpha}}_{\overline{\beta}} \zeta^{\overline{\beta}} &
\delta_\JCS \zeta^{\alpha} & \ = \ -i \tilde{\eps} {\JCS^{\alpha}}_{\overline{\beta}} \dot{\overline{Z}}^{\overline{\beta}} - \tilde{\eps} \Gamma^{\alpha}_{\beta\gamma} {\JCS^{\beta}}_{\overline{\delta}} \zeta^{\overline{\delta}} \zeta^{\gamma} \nonumber \\
\delta_\JCS \overline{Z}^{\overline{\alpha}} & \ = \ \tilde{\eps} {\JCS^{\overline{\alpha}}}_{\beta} \zeta^{\beta} &
\delta_\JCS \zeta^{\overline{\alpha}} & \ = \ -i \tilde{\eps} {\JCS^{\overline{\alpha}}}_{\beta} \dot{Z}^{\beta} - \tilde{\eps} \Gamma^{\overline{\alpha}}_{\overline{\beta}\overline{\gamma}} {\JCS^{\overline{\beta}}}_{\delta} \zeta^{\delta} \zeta^{\overline{\gamma}} \anumber \\
\delta_\KCS Z^{\alpha} & \ = \ \tilde{\eps} {\KCS^{\alpha}}_{\overline{\beta}} \zeta^{\overline{\beta}} &
\delta_\KCS \zeta^{\alpha} & \ = \ -i \tilde{\eps} {\KCS^{\alpha}}_{\overline{\beta}} \dot{\overline{Z}}^{\overline{\beta}} - \tilde{\eps} \Gamma^{\alpha}_{\beta\gamma} {\KCS^{\beta}}_{\overline{\delta}} \zeta^{\overline{\delta}} \zeta^{\gamma} \nonumber \\
\delta_\KCS \overline{Z}^{\overline{\alpha}} & \ = \ \tilde{\eps} {\KCS^{\overline{\alpha}}}_{\beta} \zeta^{\beta} &
\delta_\KCS \zeta^{\overline{\alpha}} & \ = \ -i \tilde{\eps} {\KCS^{\overline{\alpha}}}_{\beta} \dot{Z}^{\beta} - \tilde{\eps} \Gamma^{\overline{\alpha}}_{\overline{\beta}\overline{\gamma}} {\KCS^{\overline{\beta}}}_{\delta} \zeta^{\delta} \zeta^{\overline{\gamma}} \anumber
\end{align}
The corresponding supercharges are
\begin{align}
Q_{\unit} & \ = \ \phantom{i} \tilde{P}_{\alpha} \zeta^{\alpha} + \phantom{i} \tilde{P}_{\overline{\alpha}} \zeta^{\overline{\alpha}}\eqip{,} &
Q_\JCS & \ = \ \tilde{P}_{\overline{\alpha}} {\JCS^{\overline{\alpha}}}_{\alpha} \zeta^{\alpha} + \tilde{P}_{\alpha} {\JCS^{\alpha}}_{\overline{\alpha}} \zeta^{\overline{\alpha}}\eqip{,} \nonumber \\
Q_\ICS & \ = \ i \tilde{P}_{\alpha} \zeta^{\alpha} - i \tilde{P}_{\overline{\alpha}} \zeta^{\overline{\alpha}}\eqip{,} &
Q_\KCS & \ = \ \tilde{P}_{\overline{\alpha}} {\KCS^{\overline{\alpha}}}_{\alpha} \zeta^{\alpha} + \tilde{P}_{\alpha} {\KCS^{\alpha}}_{\overline{\alpha}} \zeta^{\overline{\alpha}}\eqip{.} \anumber
\end{align}
They generate the supersymmetry transformations via Dirac brackets, and they obey the $N=4$ supersymmetry algebra. The supersymmetry transformations square again to $i$ times a time-derivative, $\delta^2 = i \partial_t$, and the Hamiltonian is once more $H_0 = -\half[i]\left\{ Q_{\unit}, Q_{\unit} \right\}_{DB} = -\half[i]\left\{ Q_{\ICS_i}, Q_{\ICS_i} \right\}_{DB}$.

It is now convenient to define new linear combinations of the supercharges by
\begin{align}
\tilde{Q}\phantom{^*} & \ = \ \phantom{-}\half[i] (Q_{\unit} + i Q_\ICS) \ = \ \phantom{-}i \tilde{P}_{\overline{\alpha}} \zeta^{\overline{\alpha}}\eqip{,} &
\tilde{Q}_\JCS & \ = \ \phantom{-}\half[i] (Q_\JCS - i Q_\KCS) \ = \ \phantom{-}i \tilde{P}_{\alpha} {\JCS^{\alpha}}_{\overline{\alpha}} \zeta^{\overline{\alpha}}\eqip{,} \nonumber \\
\tilde{Q}^* & \ = \ -\half[i] (Q_{\unit} - i Q_\ICS) \ = \ -i \tilde{P}_\alpha \zeta^{\alpha}\eqip{,} &
\tilde{Q}_\JCS^* & \ = \ -\half[i] (Q_\JCS + i Q_\KCS) \ = \ -i \tilde{P}_{\overline{\alpha}} {\JCS^{\overline{\alpha}}}_{\alpha} \zeta^{\alpha}\eqip{.} \anumber
\end{align}
The only non-vanishing brackets of these supercharges are
\begin{align}
\left\{ \tilde{Q} , \tilde{Q}^* \right\}_{DB} \ = \ \left\{ \tilde{Q}_\JCS , \tilde{Q}_\JCS^* \right\}_{DB} & \ = \ i H_{\mbox{eff}}\eqip{.} \anumber
\end{align}
The supercharges $\tilde{Q}$ and $\tilde{Q}^*$ are quantised, using the quantisation procedures given above, as
\begin{align}
  \tilde{Q}\phantom{^*} & \ = \ \phantom{-}i \tilde{P}_{\overline{\alpha}} \zeta^{\overline{\alpha}} & \mapsto && \ofc^{\overline{\alpha}} \wedge \nabla_{\overline{\alpha}} & \ = \ \overline{\partial}\eqip{,} \anumber \\
  \tilde{Q}^* & \ = \ -i \tilde{P}_\alpha \zeta^{\alpha} & \mapsto && - \iota(\ofc^{\alpha}) \nabla_{\alpha} & \ = \ \overline{\partial}^{\dag}\eqip{,} \anumber
\end{align}
where $\overline{\partial}$ and $\overline{\partial}^{\dag}$ are the Dolbeault operator and its adjoint operator respectively.

For the remaining supercharges, we find that they are quantised as
\begin{align}
  \tilde{Q}_\JCS & \ = \ \phantom{-}i \tilde{P}_{\alpha} {\JCS^{\alpha}}_{\overline{\alpha}} \zeta^{\overline{\alpha}}  & \mapsto & \JCS(\ofc^{\alpha}) \ \wedge \nabla_{\alpha}={\JCS^{\alpha}}_{\overline{\alpha}} \, \ofc^{\overline{\alpha}} \wedge \nabla_{\alpha} & \ = \ \phantom{^{\dag}} \JCS \partial \JCS^{-1} \ = \ \overline{\partial}_{\JCS}\eqip{,} \label{twistdou} 
\\
  \tilde{Q}_\JCS^* & \ = \ -i \tilde{P}_{\overline{\alpha}} {\JCS^{\overline{\alpha}}}_{\alpha} \zeta^{\alpha} & \mapsto & - \iota\left(\JCS (\ofc^{\overline{\alpha}})\right) \wedge \nabla_{\overline{\alpha}} & \ = \ \JCS \partial^{\dag} \JCS^{-1} \ = \ \overline{\partial}_{\JCS}^{\dag}\eqip{,} \anumber
\end{align}
where $\overline{\partial}_{\JCS}$ and $\overline{\partial}_{\JCS}^{\dag}$ are the twisted Dolbeault operator and its adjoint respectively.

The quantisation of the Hamiltonian as the Dirac bracket of the supercharges gives once more, of course, half the Laplacian\mycite{Witten:ConstrSusyBr}:
\begin{align}
H_0 & \ = \ -i \left\{ \tilde{Q}, \tilde{Q}^* \right\}_{DB} & \mapsto &&
\overline{\partial} \, \overline{\partial}^{\dag} + \overline{\partial}^{\dag} \overline{\partial} & \ = \ \half \Delta\eqip{.} \anumber
\end{align}
This agrees with the well-known fact that under the equivalence between anti-holomorphic forms and spinors\mycite{Hitchin:HarmSpin} $\slD = \sqrt{2} (\overline{\partial} + \overline{\partial}^{\dag})$.

\subsection{The action of the complex structures}
The complex structure $\ICS$ acts on anti-holomorphic one-forms by multiplication with $i$,  but the complex structures $\JCS$ and $\KCS$ map  forms which are anti-holomorphic with respect to $\ICS$  to holomorphic forms.  This should be contrasted with the  maps $\hat{\ICS}$, $\hat{\JCS}$ and $\hat{\KCS}$ acting on the spinor zero-modes in the original field theory \eqref{actionCompStructFermZeromode}.  We can  implement these maps  on the anti-holomorphic forms on the moduli space, using their relation  to $\ICS$, $\JCS$ and $\KCS$ given in equations \eqref{actionCSFZF}, $\hat{\ICS} = \ICS$, $\hat{\JCS} = -\overline{\JCS}$ and $\hat{\KCS} = - \overline{\KCS}$. This way we obtain again an (anti-linear) action of the quaternion algebra on the space of anti-holomorphic forms, this time on the moduli space.


\section{Angular momentum and the spin operator}
\label{ansect}

The spin operator for the fermionic zero-modes of the $N=2$ supersymmetric $SU(2)$ monopole of charge 1, derived from the field theory by Osborn\mycite{Osborn:TopChSpin}, is given by
\begin{align}
\vec{S} & \ = \ \half \sum_{s,s'} a_s^{\dag} (\vec{\sigma})_{ss'} a_{s'}\eqip{,} \label{spinOpFT}
\end{align}
where $s$ and $s'$ indicate the spin state of the fermion zero-mode (ie. up or down). It acts on the Hilbert space generated by the $a_s^{\dag}$ acting on a vacuum state \ket{0}, defined by $a_s \ket{0} = 0$. The states can be grouped into two singlets and a doublet. We would like to translate Osborn's result into geometrical language, where we describe the fermionic zero-modes with anti-holomorphic forms on moduli spaces.

In general, the spin operator for higher charge monopoles is not well defined since one cannot unambiguously separate orbital angular momentum from spin due to the extended nature of monopoles. Therefore, we have to look at the total angular momentum operator $\vec{\Jang}$ instead.
However, in special cases (in particular for the charge-1 monopole, and for well separated monopoles) we expect that a spin operator, corresponding to $\vec{S}$ in equation \eqref{spinOpFT}, could reappear. This would allow us to determine the spin of individual (well separated) monopoles.
We will explicitly construct a spin-operator for the charge-1 monopole in section~\ref{sectEx1Spin}, and confirm that it agrees with Osborn's spin operator \eqref{spinOpFT}, and discuss the issues involved in computing spins of well-separated monopoles in our outlook at the end of this paper.

\subsection{The total angular momentum operator on the moduli space}
\label{sectAngMomOp}
The total angular momentum operator $\vec{\Jang}$ should act on quantum states by an infinitesimal  $SU(2)$ or $SO(3)$ action. It obeys the angular momentum algebra
\begin{align}
\left[ \Jang_i, \Jang_j \right] & \ = \ i \, \epsilon_{ijk} \Jang_k\eqip{,} \label{angMomAlgebra}
\end{align}
and acts via the Leibniz rule on tensor products of states. We expect $\vec{\Jang}$ to contain
orbital and spin contributions, but, as explained above, neither of these needs to be separately
well-defined.
It would be natural to guess that $\Jang_i$ acts by rotating the spatial coordinates through the Lie derivative $\mathcal{L}_{\sogen_i}$, where the vector fields $\sogen_i$ generate such rotations\footnote{We choose the vector fields to satisfy $\left[\sogen_i, \sogen_j\right] = \varepsilon_{ijk} \sogen_k$.}. However, we want the expression for $\vec{\Jang}$ to respect our decomposition of vectors and forms into their holomorphic and anti-holomorphic parts, because only the $(0,p)$-forms are identified with fermionic zero-modes. Therefore we require that\footnote{Here we use the adjoint action of the complex structure, which is the extension of the complex structure that acts on $p$-forms following the Leibniz rule\mycite{Verbitsky:HypHolBundHKM}.}
\begin{align}
[\Jang_i, \ad \ICS_j] & \ = \ 0\eqip{.} \label{angMomCompStr}
\end{align}
This means that the Lie derivative by itself cannot be the angular momentum operator, since Atiyah and Hitchin\mycite{AtiyahHitchin} have found that the complex structures are rotated into each other through the $SO(3)$ action: $\mathcal{L}_{\sogen_i}(\ICS_j) = \varepsilon_{ijk} \ICS_k$. For the action on $p$-forms, this becomes
\begin{align}
\left[\mathcal{L}_{\sogen_i}, \ad \ICS_j\right] & \ = \ \varepsilon_{ijk} \ad \ICS_k\eqip{.} \label{commLieadI}
\end{align}
To correct for this unwanted rotation of the complex structures, we note that the action of the complex structures generate an $SU(2)$ action on the bundle of forms themselves\mycite{Verbitsky:HypHolBundHKM, Verbitsky:QDolbCompl},
\begin{align}
\left[\ad \ICS_i, \ad \ICS_j\right] & \ = \ 2 \varepsilon_{ijk} \ad \ICS_k\eqip{,} \anumber
\end{align}
and therefore we define the operator $\Jang_i$ by
\begin{align}
\Jang_i & \ = \ i \left( \mathcal{L}_{\sogen_i} - \half \ad \ICS_i \right)\eqip{.} \label{Jang}
\end{align}
It obeys the Leibniz rule, it obeys the angular momentum algebra (equation \eqref{angMomAlgebra}), and it leaves the complex structures invariant (equation \eqref{angMomCompStr}).
In particular, the three generators $\Jang_i$ map anti-holomorphic forms to anti-holomorphic forms, even though $\Jang_1$ and $\Jang_2$ are made up of two operators which, individually,  map anti-holomorphic forms to holomorphic forms.

\subsection{Supercharges and the angular momentum operator}
\label{sectSupChAndJ}
The supercharges can be used to create spin states from bosonic states, and therefore they correspond to spin-$\half$ operators. As such, the angular momentum operators must obey the appropriate algebra with the supercharges. We find that they do, and that the Dolbeault operator $\overline{\partial}$ increases the total angular momentum of states (with respect to $\Jang_3$) by $\half$, while $\overline{\partial}_{\JCS}$ decreases it by $\half$. The supercharges are therefore indeed spin-$\half$ operators.

To derive the commutators of the $\Jang_i$ with the Dolbeault operators, we first need to compute some basic identities. First of all, we write de Dolbeault operators in terms of twisted exterior derivatives\mycite{Verbitsky:HypHolBundHKM, Verbitsky:QDolbCompl}
\begin{align}
  \overline{\partial} & \ = \ \half ( \d - i \d_\ICS )\eqip{,} &
  \overline{\partial}_{\JCS} & \ = \ \half ( \d_\JCS - i \d_\KCS )\eqip{,} \label{DolbeaultExtDer}
\end{align}
where the twisted exterior derivative\footnote{Viewing the moduli space as a K\"ahler manifold $(\MS_{\topk}, \ICS)$ we have $d_\ICS = \ICS \d \ICS^{-1} = d^c$ in conventional notation.} is defined by $\d_{\ICS_i} = {\ICS_i} \d {\ICS_i}^{-1}$. The twisted exterior derivatives can be obtained from the ordinary exterior derivative using the adjoint action of the complex structures.
\begin{align}
  \left[\ad \ICS_i, \d \right] \ = \ \left[ \ad \ICS_i, \partial_i + \overline{\partial}_i \right] \ = \ -i \left( \partial_i - \overline{\partial}_i \right) \ = \ \d_{\ICS_i} \nosum . \label{bracAdIketd}
\end{align}
Here $\partial_i$ and $\overline{\partial}_i$ are the Dolbeault operators corresponding to the complex structure $\ICS_i$ (and not to be confused with twisted Dolbeault operators).

Next we note that
\begin{align}
  \left[\mathcal{L}_{\sogen_i}, \ICS_j\right] & \ = \ \begin{cases}0 & \mbox{if } i=j \\ \ad \ICS_i \; \ICS_j & \mbox{if } i\neq j \end{cases} &
  \left[ \ad \ICS_i, \ICS_j \right] & \ = \ \begin{cases}0 & \mbox{if } i=j \\ 2 \ad \ICS_i \;\ICS_j & \mbox{if } i\neq j \end{cases} \anumber \\
\intertext{from which we compute, using \eqref{bracAdIketd}, } 
  \left[ \mathcal{L}_{\sogen_i} , \d_{\ICS_j} \right] & \ = \ \eps_{ijk} \d_{\ICS_k}\eqip{,} &
  \left[\ad \ICS_i, \d_{\ICS_j} \right] & \ = \ -\delta_{ij} \d + \eps_{ijk} \d_{\ICS_k}\eqip{.} \anumber
\end{align}
The latter equation, together with equation \eqref{bracAdIketd}, suggest that we can think of the commutator with the adjoint action of a complex structure as a twisting operator for exterior derivatives on a hyperk\"ahler manifold: when $\mathfrak{d}$ is either $\d$ or $\d_{\ICS_i}$,
\begin{align}
  \left[ \ad \ICS_i, \mathfrak{d} \right] & \ = \ \ICS_i \mathfrak{d} \ICS_i^{-1}\eqip{.} \anumber
\end{align}
Since the Dolbeault operators can be written in terms of $\d$ and $\d_{\ICS_i}$ (as in equation \eqref{DolbeaultExtDer}), this relation holds for all Dolbeault operators $\mathfrak{d}$ as well.

Finally, we define raising and lowering operators as usual by
\begin{align}
\Jang_{\pm} & \ = \ \Jang_1 \pm i \Jang_2\eqip{.}
\end{align}
The algebra satisfied by the angular momentum operators and the Dolbeault operators is then the following.
\begin{align}
  \left[ \Jang_3 , \overline{\partial} \right] & \ = \ \half \overline{\partial} &
  \left[ \Jang_3 , \overline{\partial}_{\JCS} \right] & \ = \ -\half \overline{\partial}_{\JCS}  \nonumber \\
  \left[ \Jang_+ , \overline{\partial} \right] & \ = \ 0 &
  \left[ \Jang_+ , \overline{\partial}_{\JCS} \right] & \ = \ i \overline{\partial} \nonumber \\
  \left[ \Jang_- , \overline{\partial} \right] & \ = \ -i \overline{\partial}_{\JCS} &
  \left[ \Jang_- , \overline{\partial}_{\JCS} \right] & \ = \ 0
\end{align}
We see that indeed the action of $\overline{\partial}$ increases the angular momentum of a state with respect to $\Jang_3$ by $\half$, whereas $\overline{\partial}_{\JCS}$ decreases it by $\half$.

\section{Example: charge-$1$ monopoles}
\label{sectEx1}
As a first example, we now consider a monopole of unit charge, in YMH-models with maximal symmetry breaking, for example $SU(2) \to U(1)$ or $SU(3) \to U(1) \times U(1)$. In the latter case, the charge-1 monopole is an embedding of the $SU(2)$ monopole in the $SU(3)$ model. We discuss the classical dynamics, and the quantum mechanics of the bosonic and $N=2$ supersymmetric monopole. We will explicitly exhibit the equivalence between quantisation in terms of spinors and quantisation in terms of anti-holomorphic forms on the moduli space. Finally we discuss the angular momentum and spin of charge-1 monopole states.

In this example, and the next, we use geometrical units, in which we have scaled the coupling constant and the vacuum expectation value of the Higgs field to unity,
\begin{align}
  e & \ = \ 1\eqip{,} &
  a & \ = \ 1\eqip{.} \anumber
\end{align}
This implies that Planck's constant, $\hbar$ is dimensionless, but in general not equal to unity.

The moduli space of a single monopole in this theory is\mycite{GauLowe:DyonsSDuality,LeeWeinbergYi:EMDSU3Mon},
\begin{align}
\MS_1 \ = \ \RR^3 \times S^1\eqip{.} \anumber
\end{align}
The factor $\RR^3$ corresponds to translation of the monopole, and the $S^1$ factor to long-range gauge transformations (of the form $g(\chi) = e^{\chi \Phi}$). The metric on $\MS_1$ is the flat metric
\begin{align}
\d s^2 \ = \ m \left( \d \vec{x}^2 \ + \ \d \chi^2 \right) \ = \ (\ofr^1)^2 + (\ofr^3)^2 + (\ofr^3)^2 + (\ofr^4)^2\eqip{,} \label{flatmetric}
\end{align}
where the mass of a monopole is $m = 4\pi$, the vector $\vec{x} = (x^1, x^2, x^3)$, and the vier-bein is defined by $\ofr^i = \sqrt{m} \, \d x^i$ and $\ofr^4 = \sqrt{m} \, \d \chi$. The range of $\chi$ is $0 \leq \chi < 2\pi$.

\subsection{Classical dynamics}
The classical motion of the monopole is determined by the geodesics on $\MS_1$. Since $\MS_1$ is flat, these are straight lines, corresponding to uniform motion through space and a constant electric charge.

The Lagrangian corresponding to the system of a single monopole is
\begin{align}
L & \ = \ \half[m] \left( \dot{\vec{x}} \cdot \dot{\vec{x}} \ + \ \dot{\chi}^2 \right) \ - \ m\eqip{.} \anumber
\end{align}
The Euler-Lagrange equations give us the conserved quantities momentum and electric charge
\begin{align}
\vec{p} & \ = \ \frac{\partial L}{\partial \dot{\vec{x}}} \ = \ m \, \dot{\vec{x}}\eqip{,} &
q & \ = \ \frac{\partial L}{\partial \dot{\chi}} \ = \ m \, \dot{\chi}\eqip{.} \anumber
\end{align}
The energy, given by the usual Legendre transformation
\begin{align}
E & \ = \ \frac{\partial L}{\partial \dot{\vec{x}}} \cdot \dot{\vec{x}} + \frac{\partial L}{\partial \dot{\chi}} \dot{\chi} - L \ = \ \half[m] \left( \dot{\vec{x}} \cdot \dot{\vec{x}} \ + \ \dot{\chi}^2 \right) \ + \ m\eqip{,} \anumber
\end{align}
is conserved, as well as the angular momentum $\vec{J} = \vec{x} \times \vec{p}$
.

\subsection{Quantum mechanics of bosonic monopoles}
\label{sectEx1QM}
The quantum mechanics of bosonic monopoles is most easily discussed using the quantisation in terms of forms (section~\ref{sectFZMF}). Bosonic monopoles are then described with wavefunctions ($0$-forms) on the moduli space. The Schr\"odinger equation is given by $i \hbar \partial_t \Psi = H_{\mbox{eff}} \Psi$. As we have found in section~\ref{sectFZMF}, the Hamiltonian is half the Laplacian (of the flat metric \eqref{flatmetric}) plus the mass of the monopole, so that the Schr\"odinger equation becomes
\begin{align}
i \hbar \partial_t \Psi & \ = \ \half[\hbar^2] \Delta_{\MS_1} \Psi + m \Psi \ = \ - \frac{\hbar^2}{2m} (\partial_i^2 + \partial_{\chi}^2) \Psi + m \Psi\eqip{.} \label{SchrEq}
\end{align}
By separation of variables, we find plane wave solutions
\begin{align}
\Psi & \ = \ f(t) F(\vec{x}, \chi)\eqip{,} &
f(t) & \ = \ A e^{-\frac{i}{\hbar} E t}\eqip{,} &
F(\vec{x}, \chi) & \ = \ B e^{\frac{i}{\hbar} \left( \vec{p} \cdot \vec{x}+ q \chi \right)}\eqip{,} \label{M1PlaneWaves}
\end{align}
for arbitrary constants $A$, $B$, $\vec{p}$, and $q$, such that the energy of the monopole is
\begin{align}
E \ = \ m + \frac{1}{2m} \left( |\vec{p}|^2 + q^2 \right) \eqip{,} \anumber
\end{align}
where $\vec{p}$ is its momentum, and $q$ its electric charge. Since the range of $\chi$ is $0 \leq \chi < 2\pi$, the electric charge $q$ is $\hbar$ times an integer.

The probabilistic interpretation of the wavefunction is the usual one: $| \Psi(\vec{x}, \chi) |^2$ is the probability density for the monopole at the point $(\vec{x}, \chi)$ in the moduli space. This means that $| \Psi(\vec{x}, \chi) |^2$ is the probability density for the Yang-Mills-Higgs fields of the original field theory in the (charge-1 monopole) configuration that corresponds to the point $(\vec{x}, \chi)$ in the moduli space.

\subsection{$N = 2$ supersymmetric monopoles}
Starting with a solution $\Psi$ to the bosonic Schr\"odinger equation, we can use supersymmetry to find other solutions. By applying the supercharges, ie. the Dolbeault operator $\overline{\partial}$, its twisted counterpart $\overline{\partial}_{\JCS}$ and their adjoints $\overline{\partial}^{\dag}$ and $\overline{\partial}_{\JCS}^{\dag}$, we may generate the other states of the supermultiplet containing $\Psi$. Since on a K\"ahler manifold they commute with the Hamiltonian, the solutions that we find by applying these operators to the wavefunction $\Psi$ also obey the Schr\"odinger equation.

BPS states of $N=2$ supersymmetric monopoles are those states which have minimal energy for given charges of the states. They correspond to short supermultiplets, which contain two spin-0 states, and one spin-$\half$ doublet.

A short multiplet for a charge-1 monopole is obtained whenever the spin-0 state $\Psi$ with which we start, is an eigenstate of the Hamiltonian. The states in a short multiplet can then be found using \emph{only} the (twisted) Dolbeault operators $\overline{\partial}$ and $\overline{\partial}_{\JCS}$. The states we find this way are $\Psi$, $\overline{\partial} \Psi$, $\overline{\partial}_{\JCS} \Psi$, and $\overline{\partial} \, \overline{\partial}_{\JCS} \Psi = - \overline{\partial}_{\JCS} \overline{\partial} \Psi$. For example, $\overline{\partial}^{\dag} \overline{\partial} \Psi = \half \Delta \Psi$ is not an independent state if $\Psi$ is an eigenstate of $\Delta$, for any eigenvalue.

In the next two sections we will explicitly exhibit the equivalence between anti-holomorphic forms and spinors on the moduli space. We obtain equivalent expressions for the Dirac operator and Hamiltonian for both descriptions of the fermionic zero-modes, and we give an explicit correspondence between forms and spinors.

\subsection{Quantisation using spinors}
\label{sectEx1QS}
We now construct the Dirac operator and Hamiltonian acting on spinors on $\mathcal{M}_1$. The Dirac operator is defined by
\begin{align}
\slD_s & \ = \ \frac{1}{\sqrt{m}} \gamma^a \partial_a\eqip{.} \anumber
\end{align}
We use the following representation for the Dirac $\gamma$-matrices,
\begin{align}
\gamma^j & \ = \ \left(
             \begin{array}{cc}
               0 & \sigma^j \\
               -\sigma^j & 0 \\
             \end{array}
           \right)\eqip{,} &
\gamma^4 & \ = \ \left(
               \begin{array}{cc}
                 0 & i \unit \\
                 i \unit & 0 \\
               \end{array}
             \right)\eqip{,} \label{gammaMatrices}
\end{align}
which satisfy $\{ \gamma^{\alpha}, \gamma^{\beta} \} = -2\delta^{\alpha\beta}$. The Dirac operator is therefore
\begin{align}
\slD_s & \ = \ \frac{1}{\sqrt{m}}
        \left(
          \begin{array}{cc}
            0 & \sigma^j \partial_j + i \partial_{\chi} \\
            -\sigma^j \partial_j + i \partial_{\chi} & 0 \\
          \end{array}
        \right)\eqip{.} \anumber
\end{align}
The Hamiltonian is then given by
\begin{align}
H_0 & \ = \ \half \slD_s^2 \ = \ -\frac{1}{2m} \left[ \partial_j^2 + \partial_{\chi}^2 \right] \unit_4 \ = \ \half \Delta_{\MS_1} \unit_4 \eqip{.} \anumber
\end{align}

\subsection{Quantisation using forms}
\label{sectEx1QF}
To identify the Dirac operator acting on forms, $\slD_{\overline{\partial}} = \sqrt{2} (\overline{\partial} + \overline{\partial}^{\dag})$, with the Dirac operator on spinors, we need to find an appropriate matrix representation of this operator. This implies that we need to look for a matrix representation of the Dolbeault operators. We do this by choosing a basis of anti-holomorphic forms, and representing a general anti-holomorphic form as a vector with respect to this basis. The Dolbeault operators map anti-holomorphic forms to anti-holomorphic forms and can therefore be represented by a matrix with respect to this basis.

K\"ahler coordinates for the moduli space $\MS_1 \cong \CC \times \CC^*$ are given by\mycite{AtiyahHitchin}
\begin{align}
z^2 & \ = \ x^1 + ix^2\eqip{,} &
z^1 & \ = \ e^{x^3 + i \chi}\eqip{.} \anumber
\end{align}
A convenient basis of holomorphic 1-forms, with respect to the complex structure corresponding to the hermitian coordinates $z^2$ and $z^1$, is
\begin{align}
\alpha_2 & \ = \ \sqrt{\half[m]} \, \d z^2 \ = \ \frac{1}{\sqrt{2}} (e^1 + i e^2)\eqip{,} &
\alpha_1 & \ = \ \sqrt{\half[m]} \, \frac{\d z^1}{z^1} \ = \ \frac{1}{\sqrt{2}} (e^3 + i e^4)\eqip{.} \anumber
\end{align}
The metric on $\MS_1$ then becomes
\begin{align}
\d s^2 & \ = \ m \left[ \left| \d z^2 \right|^2 + \left| \frac{\d z^1}{z^1} \right|^2 \right]
 \ = \ 2 |\alpha_2|^2 + 2 |\alpha_1|^2\eqip{.} \anumber
\end{align}

The exterior derivative is decomposed into the Dolbeault operator and its complement, $\d = \partial + \overline{\partial}$, as usual. The action of the Dolbeault operator is given by the action of the exterior derivative followed by a projection onto the anti-holomophic forms,
\begin{align}
  \overline{\partial} & \ = \ \pi^{0,\bullet} \circ \d\eqip{,} &
  \pi^{0,\bullet} : \Omega^{\bullet} (\MS_1) & \ \to \ \Omega^{0,\bullet} (\MS_1)\eqip{.} \anumber
\end{align}
We choose $\{ \overline{\alpha}_1, \overline{\alpha}_2, 1, \overline{\alpha}_1 \wedge \overline{\alpha}_2\}$ as an ordered basis of $\Omega^{0,\bullet} (\MS_1)$, to represent the action of the Dolbeault operator $\overline{\partial}$, and its adjoint $\overline{\partial}^{\dag} = - \star \partial \star$, as matrices.
\begin{align}
\big( \overline{\partial} \big) & \ = \ \left(
                           \begin{array}{cccc}
                             0 & 0 & \overline{f}_1 & 0 \\
                             0 & 0 & \overline{f}_2 & 0 \\
                             0 & 0 & 0 & 0 \\
                             - \overline{f}_2 & \overline{f}_1 & 0 & 0 \\
                           \end{array}
                         \right) &
\big( \overline{\partial}^{\dag} \big) & \ = \ \left(
                           \begin{array}{cccc}
                             0 & 0 & 0 & f_2 \\
                             0 & 0 & 0 & -f_1 \\
                             -f_1 & -f_2 & 0 & 0 \\
                             0 & 0 & 0 & 0 \\
                           \end{array}
                         \right) \anumber
\end{align}
The operators $f_1$ and $f_2$ are given by
\begin{align}
  f_1 & \ = \ \frac{1}{\sqrt{2m}} \left( \partial_3 - i \partial_{\chi} \right)\eqip{,} &
  f_2 & \ = \ \frac{1}{\sqrt{2m}} \left( \partial_1 - i \partial_2 \right)\eqip{.} \anumber
\end{align}
The Dirac operator in the chosen basis of anti-holomorphic forms is therefore the same as the Dirac operator acting on spinors, found in section~\ref{sectEx1QS},
\begin{align}
\slD_{\overline{\partial}} \ = \ \sqrt{2} \left( \overline{\partial} + \overline{\partial}^{\dag} \right) & \ = \ \frac{1}{\sqrt{m}}
     \left(
       \begin{array}{cc}
         0 & i \partial_{\chi} + \pauli_k \partial_k \\
         i \partial_{\chi} - \pauli_k \partial_k & 0 \\
       \end{array}
     \right) \ = \ \slD_s\eqip{,} \anumber
\end{align}
which means that our chosen basis of anti-holomorphic forms provides an easy way to translate between spinors and anti-holomorphic forms on the moduli space:
\begin{align}
\overline{\alpha}_1 & \ \sim \ \left(
  \begin{array}{c}
    1 \\
    0 \\
    0 \\
    0 \\
  \end{array}
\right)\eqip{,} &
\overline{\alpha}_2 & \ \sim \ \left(
  \begin{array}{c}
    0 \\
    1 \\
    0 \\
    0 \\
  \end{array}
\right)\eqip{,} &
1 & \ \sim \ \left(
  \begin{array}{c}
    0 \\
    0 \\
    1 \\
    0 \\
  \end{array}
\right)\eqip{,} &
\overline{\alpha}_1 \wedge \overline{\alpha}_2 & \ \sim \ \left(
  \begin{array}{c}
    0 \\
    0 \\
    0 \\
    1 \\
  \end{array}
\right)\eqip{.} \label{formsSpinors}
\end{align}

The Hamiltonian $H_0 = \half \slD_{\overline{\partial}}^2 = \overline{\partial}^{\dag} \overline{\partial} + \overline{\partial} \, \overline{\partial}^{\dag} = \half \Delta_{\MS_1}$ is,  of course,  also the same as the one found in section~\ref{sectEx1QS}. Since $H_0$ acts diagonally, we see that the spectrum of the $N=2$ supersymmetric monopole is simply a four-fold degenerate copy of the bosonic monopole spectrum.

We compute the matrix representation of the twisted Dolbeault operator and its adjoint,
\begin{align}
\big( \overline{\partial}_{\JCS} \big) & \ = \ i \left(
            \begin{array}{cccc}
              0 & 0 & f_2 & 0 \\
              0 & 0 & -f_1 & 0 \\
              0 & 0 & 0 & 0 \\
              f_1 & f_2 & 0 & 0 \\
            \end{array}
          \right)\eqip{,} &
\big( \overline{\partial}_{\JCS}^{\dag} \big) & \ = \ i \left(
            \begin{array}{cccc}
              0 & 0 & 0 & \overline{f}_1 \\
              0 & 0 & 0 & \overline{f}_2 \\
              \overline{f}_2 & -\overline{f}_1 & 0 & 0 \\
              0 & 0 & 0 & 0 \\
            \end{array}
          \right)\eqip{,} \anumber
\end{align}
which allows us to compute the operators $\overline{\partial} \, \overline{\partial}_{\JCS} = - \overline{\partial}_{\JCS} \overline{\partial}$ and $- \overline{\partial}^{\dag} \overline{\partial}_{\JCS}^{\dag} = \overline{\partial}_{\JCS}^{\dag} \overline{\partial}^{\dag}$.
\begin{align}
\big( \overline{\partial} \, \overline{\partial}_{\JCS} \big) & \ = \ \half[i] \left(
            \begin{array}{cccc}
              0 & 0 & 0 & 0 \\
              0 & 0 & 0 & 0 \\
              0 & 0 & 0 & 0 \\
              0 & 0 & \Delta_{\MS_1} & 0 \\
            \end{array}
          \right) &
\big( \overline{\partial}^{\dag} \overline{\partial}_{\JCS}^{\dag} \big) & \ = \ \half[i] \left(
            \begin{array}{cccc}
              0 & 0 & 0 & 0 \\
              0 & 0 & 0 & 0 \\
              0 & 0 & 0 & \Delta_{\MS_1} \\
              0 & 0 & 0 & 0 \\
            \end{array}
          \right) \anumber
\end{align}
These expressions can easily be verified using the Kodaira relations\mycite{Verbitsky:QDolbCompl} for $\overline{\partial}$ and $\overline{\partial}_{\JCS}$,
\begin{align}
  \left[ L_{\overline{\Omega}} , \overline{\partial}^{\dag} \right] & \ = \ \overline{\partial}_{\JCS}\eqip{,} &
  \left[ L_{\overline{\Omega}} , \overline{\partial}_{\JCS}^{\dag} \right] & \ = \ - \overline{\partial}\eqip{.} \label{KodairaRelations}
\end{align}
Here $L_{\overline{\Omega}}$ is an operator of exterior multiplication by $\overline{\Omega}$, and \begin{align}
  \Omega \ = \ \omega_{\JCS} + i \omega_{\KCS} \ = \ i \alpha_1 \wedge \alpha_2 \anumber
\end{align} is the canonical holomorphic symplectic form.

\subsection{Angular momentum and spin}
\label{sectEx1Spin}
The total angular momentum operator $\vec{\Jang}$ is defined by equation \eqref{Jang}, where the vector fields $\sogen_i$ generating the $SO(3)$ action are defined by
\begin{align}
\sogen_i & \ = \ -\varepsilon_{ijk} x^j \partial_k\eqip{,} &
\left[\sogen_i, \sogen_j\right] & \ = \ \varepsilon_{ijk} \sogen_k\eqip{.} \anumber
\end{align}
The Lie derivatives with respect to these vector fields obey indeed the commutation relations \eqref{commLieadI} with the complex structures.


%
We can write the Lie-derivative in terms of the exterior derivative and interior product as
\begin{align}
\mathcal{L}_{\sogen_i} & \ = \ \iota_{\sogen_i} \d + \d \iota_{\sogen_i}\eqip{.} \anumber
\end{align}
The complex structure $\ad \ICS_i$ and the term $\d \iota_{\sogen_i}$ act trivially on functions, as $0$. Furthermore, since the $(0,1)$-forms $\overline{\alpha}_1$ and $\overline{\alpha}_2$ are closed, the term $\iota_{\sogen_i} \d$ acts on them as $0$. This suggests that we identify the orbital angular momentum and spin operators as
\begin{align}
L_i & \ = \ i \left( \iota_{\sogen_i} \d \right)\eqip{,} &
S_i & \ = \ i \left( \d \iota_{\sogen_i} - \half \ad \ICS_i \right)\eqip{,} \label{defLSop}
\end{align}
which both act on forms obeying the Leibniz rule.
With these definitions,
\begin{align}
  \vec{L}(\overline{\alpha}_m) \ = \ & 0\eqip{,} &
  \vec{S}(f) \ = \ & 0\eqip{,} \anumber
\end{align}
while the spin operator acts on $\overline{\alpha}_1$ and $\overline{\alpha}_2$ as
\begin{align}
S_i(\overline{\alpha}_m) & \ = \ \half (\pauli_i)_{mn} \overline{\alpha}_n\eqip{,} \anumber
\end{align}
in perfect agreement with Osborn's result, equation \eqref{spinOpFT}.
Note in particular that the pair $\left\{\overline{\alpha}_1, \overline{\alpha}_2\right\}$ form a two-dimensional representation of $SU(2)$.


Writing a general state
in the basis $\{ \overline{\alpha}_1, \overline{\alpha}_2, 1, \overline{\alpha}_1 \wedge \overline{\alpha}_2\}$ of anti-holomorphic forms, the orbital angular momentum and spin operators take the form
\begin{align}
  \big( L_i \big) & \ = \ i \sogen_i \, \unit_4\eqip{,} &
  \big( S_i \big) & \ = \ \half \left(
                            \begin{array}{cc}
                              \sigma_i & 0 \\
                              0 & 0 \\
                            \end{array}
                          \right)\eqip{,} \anumber
\end{align}
where the vectors $\sogen_i$ act as derivatives on 
functions
, and $0$ is the $(2\times2)$-matrix of zeros.

%
%
%


\section{Example: charge-$(1,1)$ monopoles}
\label{sectEx2}
As a second example we study the system of charge-$(1,1)$ monopoles in YMH theory with
symmetry breaking $SU(3) \to U(1) \times U(1)$. These monopoles may be thought of as being composed of two constituent monopoles which are each $SU(2)$ BPS monopoles of charge 1 but embedded into $SU(2)$ subgroups associated with different simple roots of $SU(3)$. The masses of the constituents depend on the direction of the vacuum expectation value of the Higgs field in the Cartan subalgebra of $SU(3)$ and are denoted $m_1$ and $m_2$ in the following.
As in the previous section we begin with the classical dynamics, then consider the quantum mechanics of the bosonic and $N=2$ supersymmetric monopole and explicitly exhibit  the equivalence between quantisation in terms of spinors and quantisation in terms of anti-holomorphic forms on the moduli space. Finally we also discuss the angular momentum and spin of charge-$(1,1)$ monopole states.

The main references for this section are the papers  by Gauntlett and Lowe\mycite{GauLowe:DyonsSDuality}, and by  Lee, Weinberg and Yi\mycite{LeeWeinbergYi:EMDSU3Mon};
they include a detailed discussion of magnetic and electric charges and  a derivation of the metric on moduli space for charge-$(1,1)$ monopoles:
\begin{align}
  \MS_{1,1} & \ = \ \RR^3 \times \frac{\RR \times \MS_{TN}}{\ZZ}\eqip{,} \label{11mopomod}
\end{align}
where $\MS_{TN}$ is the 4-dimensional Taub-NUT manifold with a positive length parameter. Topologically $\MS_{TN} \cong \RR^4$, but it has a curved metric, given below.
For practical calculations it is usually convenient to work with the covering space of the moduli space, $\widetilde{\MS}_{1,1} = \RR^3 \times \RR \times \MS_{TN}$, and impose the  division by $\ZZ$  on the results at the end. The physics behind the $\ZZ$ action  is explained carefully in\mycite{GauLowe:DyonsSDuality,LeeWeinbergYi:EMDSU3Mon}, and can be summarised as follows. Each
of the constituent monopoles that make up a given (1,1) monopole are invariant under one of the residual $U(1)$ gauge symmetries. Thus one may pick generators of  $U(1)\times U(1)$ so that each
monopole  can carry electric charge with respect to one but not the other generator. The angular coordinates conjugate to those charges have the usual range $[0,2\pi)$, but angular coordinates
appearing in the above decomposition are related to those angles by linear transformations which
depend on the masses of the constituent monopoles and therefore have a non-standard range, which
we will specify below.

The metric on the centre of mass moduli space $\RR^3 \times \RR$ is the flat metric. Using centre of mass coordinates $\vec{R}$ and $\chi$, the metric is analogous to \eqref{flatmetric}, replacing $m$ with the total mass  $M$ of the charge-$(1,1)$ monopole system.


The metric on the Taub-NUT manifold $\MS_{TN}$ is given by
\begin{align}
\d s^2 & \ = \ \mu \Big[ V ( \d \vec{r} \cdot \d \vec{r} ) + V^{-1} (\eta_3)^2 \Big] \ = \ \mu \Big[ V \left(\d r^2 + r^2 \left( (\eta_1)^2 + (\eta_2)^2 \right) \right) + V^{-1} (\eta_3)^2 \Big]\eqip{,} \label{taubnutmetric}
\end{align}
where $\mu$ is the reduced mass of the monopole system,
\begin{align}
 V \ = \ & \left(1 + \frac{1}{r}\right)\eqip{,}
\end{align}
and
\begin{align}
\eta_1 & \ = \ - \sin \psi \d \theta + \cos \psi \sin \theta \d \phi\eqip{,} \nonumber \\
\eta_2 & \ = \ \cos \psi \d \theta + \sin \psi \sin \theta \d \phi\eqip{,} \nonumber \\
\eta_3 & \ = \ \d \psi + \cos \theta \d \phi \ = \ \d \psi + \vec{A} \cdot \d \vec{r}\eqip{,} \anumber
\end{align}
which satisfy $\d \eta_i = \half \eps_{ijk} \eta_j \wedge \eta_k$. The coordinates $\vec{r} = (x, y, z)$ correspond to the relative position of the two monopoles. We define spherical coordinates as usual by
$x = r \sin\theta \cos\phi$,
$y = r \sin\theta \sin\phi$, and
$z = r \cos\theta$.
The Euler angles $\theta$, $\phi$ and $\psi$ are coordinates on $S^3$ with the usual  ranges: $0 \leq \theta < \pi$, $0 \leq \phi < 2\pi$ and $0 \leq \psi < 4\pi$. As explained in\mycite{GauLowe:DyonsSDuality,LeeWeinbergYi:EMDSU3Mon}, the angle $\psi$ is the conjugate variable to half the difference between the electric charges of the constituent monopoles. The range $[0,4\pi)$
reflects the fact that half the difference necessarily is an element of $\frac 1 2 \ZZ$.
Finally, the division by $\ZZ$
 on the total moduli space corresponds to identifying the points
\begin{align}
(\vec{R}, \chi, \vec{r}, \psi) & \ \sim \ (\vec{R}, \chi+2\pi, \vec{r}, \psi+\tfrac{4m_2}{m_1+m_2}\pi)\eqip{,} \label{Z2identificationM11}
\end{align}
which, as explained above,  depends on the consituent monopoles'
masses\mycite{GauLowe:DyonsSDuality,LeeWeinbergYi:EMDSU3Mon}.


A similar and closely related system is that of charge-$2$ monopoles in Yang-Mills-Higgs theory
with $SU(2)$ broken to $U(1)$. In that case the moduli space is $\MS_{2} = \RR^3 \times (S^1 \times \MS_{AH})/\ZZ_2$, where the relative moduli space $\MS_{AH}$ is the 4-dimensional Atiyah-Hitchin manifold. Asymptotically, the metric of the Atiyah-Hitchin manifold approaches the Taub-NUT metric, but with opposite sign for the mass parameter, given by equation \eqref{taubnutmetric} with $V = (1 - \frac{1}{r})$. Gibbons and Manton\mycite{GibbonsManton:CQDynBPSMon} have studied the classical and quantum mechanics of this system, and to a large extent we follow their approach. We will see that the opposite sign for the mass parameter in the Taub-NUT metric for charge-$(1,1)$ monopoles has a crucial effect on the existence of bound states in this system.

\subsection{Classical dynamics}
\label{sectClassical}

Using the product structure of the moduli space, we separate the centre of mass motion and the relative motion. The centre of mass dynamics, corresponding to motion in $\RR^3 \times \RR$, is analogous to  the single monopole dynamics discussed in example 1. For the remaining part of this section, we will focus on the relative motion of the two monopoles, described by the moduli space $\MS_{TN}$.

The Lagrangian for the relative motion is
\begin{align}
L & \ = \ \half[\mu] \left[ V \left(\dot{\vec{r}} \cdot \dot{\vec{r}} \right) + V^{-1} \left(\dot{\psi} + \cos \theta \dot{\phi} \right)^2 \right]\eqip{.} \anumber
\end{align}
The Euler-Lagrange equation $\frac{\partial L}{\partial \psi} = \partial_t \frac{\partial L}{\partial \dot{\psi}}$ gives us the conserved quantity
\begin{align}
q & \ = \ \mu V^{-1} \left(\dot{\psi} + \cos \theta \dot{\phi} \right)\eqip{.}
\end{align}
The energy is given by
\begin{align}
E & \ = \ \frac{\partial L}{\partial \dot{\vec{r}}} \cdot \dot{\vec{r}} + \frac{\partial L}{\partial \dot{\psi}} \dot{\psi} - L
  \ = \ \half[\mu] \, V \left(\dot{\vec{r}} \cdot \dot{\vec{r}} + \left( \frac{q}{\mu} \right)^2 \right)\eqip{.}
\end{align}
Following Gibbons and Manton, we now define $\vec{p}$ by
\begin{align}
\frac{\partial L}{\partial \dot{\vec{r}}} & \ = \ \mu V \dot{\vec{r}} + q\vec{A} \ = \ \vec{p} + q\vec{A}\eqip{,} &
\vec{p} & \ = \ \mu V \dot{\vec{r}}\eqip{,} \anumber
\end{align}
which is only part of the momentum canonically conjugate to $\vec{r}$. The remaining equations of motion are then found to be
\begin{align}
\dot{\vec{p}} & \ = \ \cGMmp \half[\mu] \frac{\vec{r}}{r^3} \left(\dot{\vec{r}} \cdot \dot{\vec{r}} - \left( \frac{q}{\mu} \right)^2 \right) - q \frac{\dot{\vec{r}} \times \vec{r}}{r^3}
\end{align}
Two conserved quantities are the angular momentum $\vec{J}$\footnote{The quantity $\vec{J}$
is the classical angular momentum for the relative motion, and should not be confused with
the operator for the total angular momentum, despite the notational clash.}
and a Runge-Lenz type vector $\vec{K}$:
\begin{align}
   \vec{J} & \ = \ \vec{r} \times \vec{p} + q \hat{r} &
   \vec{K} & \ = \ \vec{p} \times \vec{J} \cGMmp \left( \mu E - q^2 \right) \hat{r} \eqip{.}\anumber
 \end{align}
Since $\vec{r} \times \vec{p}$ and $\hat{r}$ are orthogonal, the magnitude of the orbital angular momentum, $l = \left| \vec{r} \times \vec{p} \right|=\sqrt{J^2-q^2}$, is also conserved.
In order to determine the orbits we need to distinguish the cases $q=0$ and $q\neq 0$.
In the former case  one checks that
\begin{align}
\label{q0law}
  \vec{J} \cdot \vec{r} & \ = \ 0\eqip{,} &
  \vec{K} \cdot \vec{r} & \ = \ J^2 - \mu E r\eqip{,} \anumber
\end{align}
as well as $\vec{J}\cdot \vec{K}=0$. Thus the motion takes place in the plane orthogonal  to $\vec{J}$,
and the vector $\vec{K}$ is contained in that plane. In terms of polar coordinates $(r,\varphi)$ in that plane, with $\varphi=0$ corresponding to the direction of $\vec{K}$, the second equation in
\eqref{q0law} becomes
\begin{align}
r=\frac{J^2}{K\cos\varphi +\mu E}.
\end{align}
Since $K=|\vec{K}|= \sqrt{2\mu E J^2 + \mu^2E^2}> \mu E$,  this is the equation for a hyperbola.
For $q \neq 0$ the orbits  are determined by the simultaneous equations
\begin{align}
  \vec{J} \cdot \hat{r} & \ = \ q\eqip{,} &
  \vec{N} \cdot \vec{r} & \ = \ J^2 - q^2\eqip{,} \anumber
\end{align}
where $\vec{N}=  \vec{K} + \frac{1}{q}(\mu E - q^2) \vec{J} $.
The first of these is the equation of a cone with axis $\vec{J}$ and  opening angle $2\alpha$
determined by $J \cos\alpha=q$. The second is the equation of a plane
orthogonal to the vector $\vec{N}$.  Hence the
orbits are  conic sections, but to  determine  which kinds of conic sections occur  we need to find  the
angle between the vectors $\vec{N}$ and $\vec{J}$. A lengthy calculation shows that
\begin{align}
\label{NJ}
N=|\vec{N}| & \  =  \  \frac{l\mu E}{|q|} \eqip{,} &
\vec{N}\cdot \vec{J}& \  =  \  \frac{l^2(\mu E-q^2)}{q}\eqip{.}
\end{align}
Let us assume for simplicity  that $q>0$ from now on,
so that  the vector $\vec{J}$ is inside the cone, and that
$\alpha \in[0,\tfrac\pi 2)$;  the case $q<0$ can be dealt with analogously, but using $-\vec{J}$
instead of $\vec{J}$    (the second equation in \eqref{NJ} is invariant under simultaneous sign change
of $q$ and $\vec{J}$).  It   then follows  from \eqref{NJ} that
the angle $\beta$ between $\vec{N}$ and $\vec{J}$ satisfies
\begin{align}
\cos\beta= \frac{\mu E-q^2}{\mu E}\frac{l}{J}\eqip{.}
\end{align}
On the other hand we can use  $q^2+l^2=J^2$ to see  that the complementary angle to half of the opening angle of the cone satisfies
\begin{align}
\cos (\tfrac \pi  2- \alpha)= \frac{l}{J}\eqip{.}
\end{align}
Hence  $\cos\beta <\cos (\tfrac \pi  2- \alpha)$ or $\beta >(\tfrac \pi  2- \alpha)$. We conclude that
the intersection of the plane and the cone is always hyperbolic and that all orbits
are unbounded.

\subsection{Quantum mechanics}
\label{sectQm}
The Hamiltonian for the charge-$(1,1)$ monopole system is given by half the Laplacian on the total moduli space $\MS_{1,1}$, plus the total mass. Separating center of mass and relative motion variables using the product structure of the moduli space, the Schr\"odinger equation becomes
\begin{align}
i \hbar \partial_t \Psi & \ = \ \half[\hbar^2] \Delta_{\MS_{1,1}} \Psi + M \Psi \ = \ \half[\hbar^2] \left( \Delta_{\RR^3 \times \RR} + \Delta_{\MS_{TN}} \right) \Psi +  M \Psi\eqip{.} \label{SchrEq2}
\end{align}
Now we separate variables by assuming
\begin{align}
\Psi(t, \vec{R}, \chi, \vec{r}, \psi) & \ = \ f(t) F(\vec{R}, \chi) \Phi(\vec{r}, \psi)\eqip{,} \anumber
\end{align}
where $F(\vec{R}, \chi)$ is a wavefunction corresponding to the centre of mass motion, and $\Phi(\vec{r}, \psi)$ is a wavefunction corresponding to the relative motion. For stationary states, the Schr\"odinger equation reduces to the following Schr\"odinger equation on the relative moduli space:
\begin{align}
\half[\hbar^2] \Delta_{\MS_{TN}} \Phi & \ = \ E \Phi\eqip{,} \label{SchrEq2a}
\end{align}
where $E$ is the energy of the relative motion. The total energy is given by the sum of the energy of the total mass, the centre of mass motion and the energy of the relative motion, $E_{\text {\small total}} = 
M + \tfrac{1}{2 M} \left(|\vec{P}|^2 + Q^2\right) + E$, where $\vec{P}$ is the total momentum and $Q$ is a kind of "centre of mass" electric charge. Since the constituent monopoles are charged with respect
to different $U(1)$ groups, the charge $Q$ is a linear combination of different kinds of electric charges\mycite{GauLowe:DyonsSDuality,LeeWeinbergYi:EMDSU3Mon}; in particular  it is not necessarily
an integer but obeys a quantisation condition that follows from \eqref{Z2identificationM11}. From now onwards we shall
assume  that for plane wave solutions of the centre of mass wavefunction $F(\vec{R}, \chi)$ analogous to \eqref{M1PlaneWaves} the value of $Q$ is such that \eqref{Z2identificationM11} holds. This does not affect the relative motion, to which we now turn.

The Laplacian on $\MS_{TN}$ can be computed from the metric on the moduli space \eqref{taubnutmetric}
\begin{align}
\Delta_{\MS_{TN}} f & \ = \ - \frac{1}{\mu} \left[ \frac{1}{r^2 V} \partial_r \left[r^2 \partial_r f \right] + \frac{1}{r^2 V} \left[\xi_1^2 + \xi_2^2\right] f + V \xi_3^2 f \right]\eqip{,} \anumber
\end{align}
where $\xi_i$ are the vector fields dual to $\eta_i$, $\eta_i (\xi_j) = \delta_{ij}$:
\begin{align}
\xi_1 & \ = \ -\frac{\cos\theta}{\sin\theta} \cos\psi \frac{\partial}{\partial \psi} - \sin\psi \frac{\partial}{\partial \theta} + \frac{\cos\psi}{\sin\theta} \frac{\partial}{\partial \phi}\eqip{,} \nonumber \\
\xi_2 & \ = \ -\frac{\cos\theta}{\sin\theta} \sin\psi \frac{\partial}{\partial \psi} + \cos\psi \frac{\partial}{\partial \theta} + \frac{\sin\psi}{\sin\theta} \frac{\partial}{\partial \phi}\eqip{,} \nonumber \\
\xi_3 & \ = \ \frac{\partial}{\partial \psi}\eqip{.} \anumber
\end{align}
Writing $\epsilon = \frac{2\mu E}{\hbar^2}$, and multiplying with $V$, the Schr\"odinger equation \eqref{SchrEq2a} becomes
\begin{align}
\frac{1}{r^2} \partial_r \left[r^2 \partial_r \Phi \right] + \frac{1}{r^2} \left[\xi_1^2 + \xi_2^2 + \xi_3^2 \right] \Phi + \left(1 \cGMpm \frac{2}{r}\right) \xi_3^2 \Phi + \epsilon \left(1 \cGMpm \frac{1}{r}\right) \Phi & \ = \ 0\eqip{.} \label{SchrEq2b}
\end{align}

\subsubsection{There are no bound states for the bosonic monopole}
We expand $\Phi$ in terms of  the Wigner functions $D^j_{sm} (\theta,\phi,\psi) =  e^{im\phi} d^j_{sm}(\theta) e^{is\psi}$  on $SU(2)$ with indices  $j,s,m\in\frac 1 2 \ZZ$ and $j\geq 0$.
They are  are eigenfunction of
$\xi_3$ and $(\xi_1^2 + \xi_2^2 + \xi_3^2)$\mycite{LandauLifshitz}:
\begin{align}
\xi_3 D^j_{sm} & \ = \ i s D^j_{sm}\eqip{,} &
(\xi_1^2 + \xi_2^2 + \xi_3^2) D^j_{sm} & \ = \ -j(j+1) D^j_{sm}\eqip{.} \anumber
\end{align}
For $\Phi = \frac{h(r)}{r} D^j_{sm}(\theta, \phi, \psi)$ the Schr\"odinger equation \eqref{SchrEq2b} reduces to
\begin{align}
\frac{1}{r} \left[ {\partial_r}^2 - \frac{j(j+1)}{r^2} + (\eps - s^2) \cGMpm \frac{1}{r} (\eps - 2 s^2) \right] h(r) & \ = \ 0\eqip{.} \label{SchrEq2c}
\end{align}
To solve this equation, we use the Ansatz $h(r) = r^{j+1} e^{ikr} F$ and find
\begin{align}
h(r) & \ = \ r^{j+1} e^{ikr} F\left( (j+1) + i\lambda, 2(j+1), -2ikr \right)\eqip{,} \anumber
\end{align}
where $F(a,b,u)$ is a confluent hypergeometric function, and we have defined
\begin{align}
k^2 & \ = \ \left( \epsilon - s^2 \right)\eqip{,} &
\lambda & \ = \ \cGMmp \frac{1}{2k} ( \epsilon - 2 s^2 ) \ = \ \cGMmp \frac{1}{2k} (k^2 - s^2)\eqip{.} \anumber
\end{align}

Bound states correspond to square integrable solutions of $h(r)$. For these, the exponential term must vanish for large $r$ and the series expansion of $F$ must terminate. For the exponential term to vanish, $k$ must be $i$ times a positive real number ($ik<0$), so
\begin{align}
\epsilon & \ < \ s^2\eqip{.} \label{condExpVanishes}
\end{align}
The expansion of $F(a,b,u)$ is
\begin{align}
F(a,b,u) & \ = \ 1 + \frac{a}{b} u + \frac{a(a+1)}{b(b+1)} \frac{u^2}{2!} + \ldots\eqip{,} \anumber
\end{align}
which terminates if $a$ is a non-positive integer. In this case $a = (j+1) + i\lambda$, so we require
\begin{align}
-i\lambda & \ = \ n, \qquad n = j+1, j+2, \ldots\eqip{.} \anumber
\end{align}
Since $-ik>0$ and $j\geq 0$, the only solutions occur when
\begin{align}
\epsilon - 2 s^2
& \ > \ 0\eqip{.} \anumber
\end{align}
which contradicts equation (\ref{condExpVanishes}). Hence there are no quantum mechanical bound states in this system, reflecting the absence of bound orbits in the corresponding classical system.

\subsubsection{Scattering states}
\label{scattstates}
To find the scattering states we follow Gibbons and Manton's approach, and introduce parabolic coordinates \begin{align}
\xi & \ = \ r + z\eqip{,} &
\eta & \ = \ r - z\eqip{.} \anumber
\end{align}
We now use the Ansatz
\begin{align}
\Phi & \ = \ e^{im\phi}e^{is\psi}\Lambda(\xi,\eta)\eqip{,} \anumber
\end{align}
with $s\in \frac 1 2 \ZZ$  in order to respect the range of $\psi$.
We find that the Schr\"odinger equation \eqref{SchrEq2b} reduces to
\begin{align}
0 & \ = \ \frac{4}{\xi+\eta} \left[ \partial_{\xi}(\xi\partial_{\xi} \Lambda) + \partial_{\eta}(\eta\partial_{\eta} \Lambda)\right] - \frac{1}{\xi\eta} \left( m^2 + s^2 - 2ms \frac{\xi - \eta}{\xi + \eta} \right) \Lambda \nonumber \\
 & \qquad + \left( \epsilon - s^2 \right) \Lambda \cGMpm \frac{2}{\xi + \eta} \left( \epsilon - 2 s^2 \right) \Lambda\eqip{.} \anumber
\end{align}
To solve this equation, we separate variables as follows,
\begin{align}
\Lambda(\xi,\eta) & \ = \ f(\xi) g(\eta)\eqip{,} \anumber
\end{align}
and we find that $f$ and $g$ must satisfy
\begin{align}
\frac{4}{f} \partial_{\xi}(\xi\partial_{\xi} f) - \frac{1}{\xi} (m + s)^2 + k^2 \xi \cGMpm 2 \left( 2\epsilon - s^2 \right) & \ = \ C\eqip{,} \label{eqf} \\
- \frac{4}{g} \partial_{\eta}(\eta\partial_{\eta} g) + \frac{1}{\eta} (m - s)^2 - k^2 \eta & \ = \ C\eqip{.} \label{eqg}
\end{align}
We use the following trial solutions:
\begin{align}
f(\xi) & \ = \ \xi^{\half |m+s|} e^{-ik\xi/2} F_1(\xi)\eqip{,} &
g(\eta) & \ = \ \eta^{\half |m-s|} e^{-ik\eta/2} F_2(\eta)\eqip{,} \anumber
\end{align}
and find that $F_1$ and $F_2$ are again confluent hypergeometric functions:
\begin{align}
F_1 & \ = \ F(c_1, |m+s| + 1, ik\xi)\eqip{,} &
F_2 & \ = \ F(c_2, |m-s| + 1, ik\eta)\eqip{,} \anumber
\end{align}
where $c_1$ and $c_2$ are constants that must satisfy
\begin{align}
c_1 + c_2 & \ = \ 1 + \half |m+s| + \half |m-s| - i\lambda\eqip{.} \anumber
\end{align}
We can use the remaining freedom to specify the scattering situation we want to describe.
The constituent monopoles are distinguishable particles (with different magnetic charges and,
in general, different masses) so we can label them 1 and 2. Then we can assume without loss
of generality that $\vec{r}$ is the position vector of monopole 1 relative to monopole 2, and $\psi$
the phase  of monopole 1 relative to that of monopole 2 (see\mycite{GauLowe:DyonsSDuality,LeeWeinbergYi:EMDSU3Mon} for details), so that $s = \frac 1 2 $ (electric
charge of monopole 1 minus electric charge of monpole 2). We would like to consider scattering where
monopole $1$ comes in along the negative $z$-axis, and monopole $2$ along the positive $z$ axis.
This can be achieved  by setting $m=-s$ and $c_1=1$:
\begin{align}
\Phi & \ = \ e^{is(\psi-\phi)} \ (r-z)^{|s|} \ e^{+ikz} \ F(|s| - i \lambda, 2|s| + 1, ik (r-z))\eqip{,} \anumber
\end{align}
where we have used the fact that $F(1,1,x) = e^x$.
To compute the scattering cross section, we will need to find the asymptotic form of 
$\Phi$. Expanding $F$, we find
\begin{align}
\Phi & \ \approx \ e^{is(\psi-\phi)} K \left\{ \left( 1 + \frac{(s^2+\lambda^2)}{2ikr\sin^2\left(\frac{\theta}{2}\right)} \right) e^{i\left(kz + \lambda \log(k(r-z))\right)} + \right. \nonumber \\
 & \qquad \qquad \qquad \qquad \left. \frac{(|s| - i \lambda)}{2ikr\sin^2\left(\frac{\theta}{2}\right)} e^{i(\tau + \pi |s|)} e^{i\left(kr - \lambda\log(k(r-z)) \right)} \right\}\eqip{,} \anumber
\end{align}
where
\begin{align}
K & \ = \ \frac{\Gamma(2|s| + 1)}{\Gamma(|s| + 1 + i \lambda)} \frac{1}{(-i)^{|s| - i \lambda} k^{|s|}}\eqip{,} &
\tau & \ = \ \mbox{arg} \frac{\Gamma(|s| + 1 + i \lambda)}{\Gamma(|s| + 1 - i \lambda)}\eqip{.} \anumber
\end{align}
The differential cross sections is thus the same as found by Gibbons and Manton\mycite{GibbonsManton:CQDynBPSMon}  in the Taub-NUT approximation to
dyon scattering in $SU(2)$  Yang-Mills-Higgs theory:
\begin{align}
\frac{\d \sigma}{\d \Omega} & \ = \ \frac{1}{4}\left(1 + \frac{s^2}{4k^2}\right)^2 \sin^{-4}\left(\frac{\theta}{2}\right)\eqip{.} \anumber
\end{align}
It is  interesting  that a very similar cross section is found for the scattering of 
a charged particle off a BPS monopole\mycite{FH} when exponential terms are neglected.

In the case where the relative electric charge $s$ vanishes, which includes
the case of pure monopole scattering, one obtains the purely geometric (energy
independent) expression
\begin{align}
\frac{\d \sigma}{\d \Omega} & \ = \ \frac{1}{4} \sin^{-4}\left(\frac{\theta}{2}\right)\eqip{.} \anumber
\end{align}
It was shown by one of the authors\mycite{Schroers:QSBPSMon} that the symmetrised version of   this formula
is also a very good approximation to the differential cross section  for  pure monpole scattering in the $SU(2)$ theory: the s-wave phase shift correction to the Taub-NUT approximation is zero for all energies in that case. Remarkably, two identical $SU(2)$ monopoles and  the two distinct $SU(3)$ monopoles  that make up the charge-$(1,1)$ configuration thus have the same scattering behaviour in the quantum theory at low energies, apart from  symmetrisation effects.


\subsection{$N = 2$ supersymmetric monopoles}
\label{sectexN=2}
As before, we use the product structure of the moduli space to separate centre of mass and relative motion variables. This procedure is most transparent using the quantisation in terms of forms on the moduli space, although the equivalence between anti-holomorphic forms and spinors assures us that it can be done for the latter as well.

Anti-holomorphic forms on the total moduli space can be written as wedge products of anti-holomorphic forms on the centre of mass and relative moduli spaces. The supercharges, the (twisted) Dolbeault operators and their adjoints, decompose into a sum of (twisted) Dolbeault operators on the centre of mass and relative moduli spaces. The Laplacian is then seen to decompose into the sum of centre of mass and relative moduli space components as well. For an anti-holomorphic form $\overline{\upsilon} = \overline{\upsilon}_{1} \wedge \overline{\upsilon}_{2}$, where $\overline{\upsilon}_{1}$ and $\overline{\upsilon}_{2}$ are anti-holomorphic forms on $\RR^3 \times \RR$ and $\MS_{TN}$ respectively,
\begin{align}
  \overline{\partial}_{M_{1,1}} \overline{\upsilon} & \ = \ ( \overline{\partial}_{\RR^3 \times \RR} \overline{\upsilon}_{1} ) \wedge \overline{\upsilon}_{2} + (-1)^{\deg(\overline{\upsilon}_{1})} \overline{\upsilon}_{1} \wedge ( \overline{\partial}_{\MS_{TN}} \overline{\upsilon}_{2} )\eqip{,} \nonumber \\
  \Delta_{\MS_{1,1}} \overline{\upsilon} & \ = \ ( \Delta_{\RR^3 \times \RR} \overline{\upsilon}_{1} ) \wedge \overline{\upsilon}_{2} + \overline{\upsilon}_{1} \wedge ( \Delta_{\MS_{TN}} \overline{\upsilon}_{2} )\eqip{.} \anumber
\end{align}

Focussing on the moduli space for the relative motion of the monopoles, we can as before generate multiplets of states, starting with a wavefunction $\Phi$ on $\MS_{TN}$, and applying the (twisted) Dolbeault operators on the Taub-NUT manifold. When $\Phi$ is an eigenstate of the Laplacian $\Delta_{\MS_{TN}}$ we obtain, in general, four independent states. By taking the wedge product of these states with a multiplet of four centre of mass states, we obtain a multiplet of 16 states with the same energy.

BPS states in the original field correspond to the short multiplet of 4 states on the total moduli space. Therefore, these would correspond to a normalisable harmonic form on the relative moduli space, which does not generate a multiplet of 4 independent states on the moduli space. However, the only harmonic form on the Taub-NUT manifold is of bidegree $(1,1)$\mycite{GauLowe:DyonsSDuality,LeeWeinbergYi:EMDSU3Mon}.  It is therefore not an anti-holomorphic state corresponding to any fermionic or bosonic state of the system. The $N=2$ supersymmetric charge-$(1,1)$ monopole system therefore has no BPS states.

\subsection{Quantisation using spinors}
\label{sectEx2QS}
Having described how we may separate the centre of mass motion and relative motion in the previous section, we now focus our attention on the relative moduli space. We construct the Dirac operator and Hamiltonian acting on spinors on $\MS_{TN}$, which we will compare to, and see to be equivalent to, the corresponding operators acting on forms in section~\ref{sectEx2QF}.

The Taub-NUT metric can be rewritten as
\begin{align}
\d s^2 & \ = \ \mu \Big[ V (\d \vec{r})^2 + V^{-1} \left( \d \psi + \vec{A} \cdot \d \vec{r} \right)^2 \Big]
 \ = \ (e^1)^2 + (e^2)^2 + (e^3)^2 + (e^4)^2\eqip{,} \anumber
\end{align}
where as before $\vec{A} \cdot \d \vec{r} = \cos \theta \d \phi$, and we have defined the vier-bein
\begin{align}
e^i & \ = \ {e^i}_{\underline{j}} \d x^{\underline{j}} \ = \ \sqrt{\mu V} \d x^i\eqip{,} &
e^4 & \ = \ {e^4}_{\underline{j}} \d x^{\underline{j}} \ = \ \sqrt{\frac{\mu}{V}} \left( \d \psi + \vec{A} \cdot \d \vec{r} \right)\eqip{.} \anumber
\end{align}

The Dirac operator on a general manifold is defined by
\begin{align}
\slD_s & \ = \ \hat{\gamma}^a \left( \partial_a + \Gamma_a \right)\eqip{,} \anumber
\end{align}
where
\begin{align}
\hat{\gamma}^a & \ = \ ({e^a}_b)^{-1} \gamma^b\eqip{,} &
\Gamma_a & \ = \ - \frac{1}{8} [\gamma^b, \gamma^c] \omega_{bc}(\partial_a)\eqip{,} \anumber
\end{align}
and the spin connection $\omega$ corresponding to the viel-bein $e$ is defined through
\begin{align}
\d e^a + {\omega^a}_b \wedge e^b & \ = \ 0\eqip{,} &
\omega_{ab} & \ = \ - \omega_{ba}\eqip{.} \label{defomega}
\end{align}
We use the same representation for the Dirac $\gamma$-matrices as before
\eqref{gammaMatrices}, so that
\begin{align}
\hat{\gamma}^j & \ = \ \frac{1}{\sqrt{\mu}} \left(
                     \begin{array}{cc}
                       0 & \frac{1}{\sqrt{V}} \sigma^j \\
                       -\frac{1}{\sqrt{V}} \sigma^j & 0 \\
                     \end{array}
                   \right)\eqip{,} \nonumber \\
\hat{\gamma}^4 & \ = \ \frac{1}{\sqrt{\mu}} \left(
                     \begin{array}{cc}
                       0 & i \sqrt{V} - \frac{1}{\sqrt{V}} \sigma^j A^j \\
                       i \sqrt{V} + \frac{1}{\sqrt{V}} \sigma^j A^j & 0 \\
                     \end{array}
                   \right)\eqip{.} \anumber
\end{align}
The Dirac operator is therefore
\begin{align}
\slD_s & \ = \ \frac{1}{\sqrt{\mu}}
        \left(
          \begin{array}{cc}
            0 & i \frac{1}{\sqrt{V}} \sigma^j \pi^j + i \sqrt{V} \partial_{\psi} \\
            -i \frac{1}{V} \sigma^j \pi^j \sqrt{V} + i \sqrt{V} \partial_{\psi} & 0 \\
          \end{array}
        \right) \ =: \ \frac{1}{\sqrt{\mu}}
        \left(
           \begin{array}{cc}
             0 & T^{\dag} \\
             T & 0 \\
           \end{array}
         \right)\eqip{,} \anumber
\end{align}
where we have defined the operator $T$ and its adjoint $T^{\dag}$, and
\begin{align}
\pi^j & \ = \ -i (\partial_j - A^j \partial_{\psi})\eqip{.} \label{piderivative}
\end{align}
The Hamiltonian is then given by
\begin{align}
H_0 & \ = \ \half \slD_s^2 \ = \
         \left(
           \begin{array}{cc}
             H_2 & 0 \\
             0 & H_1 \\
           \end{array}
         \right)\eqip{,} \label{Ex2HS}
\end{align}
where
\begin{align}
H_1 & \ = \ \frac{1}{2\mu} T T^{\dag} \ = \ \frac{1}{2\mu} \left[ \frac{1}{V} \pi^j \pi^j - V \partial_{\psi}^2 \right]\eqip{,} \anumber \\
H_2 & \ = \ \frac{1}{2\mu} T^{\dag} T \ = \ H_1 + \frac{1}{2\mu} \left[ \frac{3}{4V^3 r^4} + \frac{1}{V^2 r^3} \vec{\sigma} \cdot \vec{L}_0 - i \frac{1}{V^2 r^3} \vec{\sigma} \cdot \vec{r} \partial_{\psi} \right]\eqip{,} \anumber
\end{align}
and
\begin{align}
\vec{L}_0 & \ = \ \vec{r} \times \vec{\pi} -i \hat{r} \partial_{\psi}\eqip{.} \anumber
\end{align}
This disagrees with the result found by Comtet and Horv\'athy\mycite{ComHor:DiracTaubNUT}, who studied the Dirac equation in Taub-NUT space in the context of gravitational instantons.
Notice that $H_1$ acts diagonally - a fact we will come to appreciate further when we compute the Dirac operator acting on forms in the following section.

\subsection{Quantisation using forms}
\label{sectEx2QF}
Once again we will construct a matrix representation of the Dirac operator action on anti-holomorphic forms, by finding a matrix representation of the Dolbeault operators with respect to a suitable basis of anti-holomorphic forms.

A set of K\"ahler coordinates on the Taub-NUT manifold $\MS_{TN}$ are defined by\mycite{GibRub:HiddenSymMM}
\begin{align}
w & \ = \ r \sin \theta e^{i\phi}
\eqip{,} &
v & \ = \ r (1+\cos\theta) e^{r \cos \theta +i(\psi+\phi)}\eqip{.} \anumber
\end{align}
We define the 1-forms $\alpha_2$ and $\alpha_1$, which form a convenient basis of holomorphic 1-forms with respect to the complex structure corresponding to the hermitian coordinates $w$ and $v$, by
\begin{align}
\alpha_2 & \ = \ \frac{1}{\sqrt{2}} (e^1 + i e^2)
  \ = \ \sqrt{\frac{\mu V}{2}} \ \d w\eqip{,} \nonumber \\
\alpha_1 & \ = \ \frac{1}{\sqrt{2}} (e^3 + i e^4)
  \ = \ \sqrt{\frac{\mu}{2V}} \left( \frac{\d v}{v} + (\cos\theta - 1) \frac{\d w}{w} \right)\eqip{.} \anumber
\end{align}
The Taub-NUT metric can then be written in terms of K\"ahler coordinates as
\begin{align}
\d s^2 & \ = \ \mu \left[ V \left| \d w \right|^2 + V^{-1} \left| \frac{\d v}{v} + (\cos \theta - 1) \frac{\d w}{w} \right|^2 \right]
 \ = \ 2 |\alpha_2|^2 + 2 |\alpha_1|^2\eqip{.} \anumber
\end{align}

Again we choose $\{\overline{\alpha}_1, \overline{\alpha}_2, 1, \overline{\alpha}_1 \wedge \overline{\alpha}_2\}$ as and ordered basis of $\Omega^{0,\bullet} (M)$, and using the same procedure as before, we represent the the action of the Dolbeault operator $\overline{\partial}$, and its adjoint $\overline{\partial}^{\dag} = - \star \partial \star$, as a matrices. A lengthy, but straightforward calculation gives the following result.
\begin{align}
\big( \overline{\partial} \big) & \ = \ \left(
                           \begin{array}{cccc}
                             0 & 0 & \overline{f}_1 & 0 \\
                             0 & 0 & \overline{f}_2 & 0 \\
                             0 & 0 & 0 & 0 \\
                             \overline{g}_1 - \overline{f}_2 & \overline{g}_2 + \overline{f}_1 & 0 & 0 \\
                           \end{array}
                         \right) &
\big( \overline{\partial}^{\dag} \big) & \ = \ \left(
                           \begin{array}{cccc}
                             0 & 0 & 0 & f_2 \\
                             0 & 0 & 0 & -f_1 \\
                             -g_2 -f_1 & g_1 -f_2 & 0 & 0 \\
                             0 & 0 & 0 & 0 \\
                           \end{array}
                         \right) \anumber
\end{align}
The functions $g_1$ and $g_2$, and operators $f_1$ and $f_2$, are given by
\begin{align}
g_1 & \ = \ - \frac{1}{\sqrt{\mu}} \, \frac{1}{2V\sqrt{2V}} \ (\partial_1 - i \partial_2) V\eqip{,} &
  f_1 & \ = \ i \frac{1}{\sqrt{\mu}} \frac{1}{\sqrt{2V}} \left( \pi^3 - V \partial_{\psi} \right)\eqip{,} \nonumber \\
g_2 & \ = \ \phantom{-} \frac{1}{\sqrt{\mu}} \, \frac{1}{2V\sqrt{2V}} \ (\partial_3 V)\eqip{,} &
  f_2 & \ = \ \phantom{i} \frac{1}{\sqrt{\mu}} \frac{1}{\sqrt{2V}} \left( \pi^2 + i \pi^1 \right)\eqip{,} \anumber
\end{align}
where $\pi^j$ was defined in \eqref{piderivative}. The Dirac operator in the chosen basis of anti-holomorphic forms is therefore the same as the Dirac operator acting on spinors, found in section~\ref{sectEx2QS}.
\begin{align}
\slD_{\overline{\partial}} \ = \ \sqrt{2} \left( \overline{\partial} + \overline{\partial}^{\dag} \right) & \ = \ \frac{1}{\sqrt{\mu}}
     \left(
       \begin{array}{cc}
         0 & i \sqrt{V}\partial_{\psi} + i \frac{1}{\sqrt{V}} \sigma^k \pi^k \\
         i \sqrt{V}\partial_{\psi} - i \frac{1}{V} \sigma^k \pi^k \sqrt{V} & 0 \\
       \end{array}
     \right) \ = \ \slD_s \anumber
\end{align}
This means that our chosen basis of anti-holomorphic forms gives an easy way to translate between spinors on the moduli space and anti-holomorphic forms on the moduli space.

The Hamiltonian $H = \half \slD_{\overline{\partial}}^2 = \left( \overline{\partial}^{\dag} \overline{\partial} + \overline{\partial} \, \overline{\partial}^{\dag} \right) = \half \Delta_{\MS_{1,1}^0}$ is,  of course,  also the same as the Hamiltonian for spinors on the moduli space, \eqref{Ex2HS}. We see that the effective Hamiltonian for the bosonic fields in the original field theory (the 0-forms and 2-forms on the moduli space) is $H_0$. Since it is diagonal, it acts on functions and 2-forms independently. The Hamiltonian for fermionic fields $H_1$, however, mixes the two different fermionic modes through the terms involving the Pauli-matrices. Since the Hamiltonian commutes with the Dolbeault operators, the spectrum for the fermionic sector of the theory must be the same as the spectrum for the bosonic sector. This agrees with Comtet and Horv\'athy's argument using supersymmetry on the moduli space to show that the bosonic and fermionic spectra are the same.

Finally, we compute the matrix representation of the twisted Dolbeault operator and its adjoint.
\begin{align}
\overline{\partial}_{\JCS} & \ = \ i \left(
            \begin{array}{cccc}
              0 & 0 & f_2 & 0 \\
              0 & 0 & -f_1 & 0 \\
              0 & 0 & 0 & 0 \\
              g_2 + f_1 & - (g_1 - f_2) & 0 & 0 \\
            \end{array}
          \right) &
\overline{\partial}_{\JCS}^{\dag} & \ = \ i \left(
            \begin{array}{cccc}
              0 & 0 & 0 & \overline{f}_1 \\
              0 & 0 & 0 & \overline{f}_2 \\
              -(\overline{g}_1 - \overline{f}_2) & -\overline{g_2}-\overline{f}_1 & 0 & 0 \\
              0 & 0 & 0 & 0 \\
            \end{array}
          \right) \anumber
\end{align}
We can again directly compute the operators $\overline{\partial} \, \overline{\partial}_{\JCS} = - \overline{\partial}_{\JCS} \overline{\partial}$ and $- \overline{\partial}^{\dag} \overline{\partial}_{\JCS}^{\dag} = \overline{\partial}_{\JCS}^{\dag} \overline{\partial}^{\dag}$, although 
it is easier to use the Kodaira relations 
\eqref{KodairaRelations}. This gives once more the simple result
\begin{align}
\overline{\partial} \, \overline{\partial}_{\JCS} & \ = \ \half[i] \left(
            \begin{array}{cccc}
              0 & 0 & 0 & 0 \\
              0 & 0 & 0 & 0 \\
              0 & 0 & 0 & 0 \\
              0 & 0 & \Delta_{\MS_{TN}} & 0 \\
            \end{array}
          \right) &
\overline{\partial}^{\dag} \overline{\partial}_{\JCS}^{\dag} & \ = \ \half[i] \left(
            \begin{array}{cccc}
              0 & 0 & 0 & 0 \\
              0 & 0 & 0 & 0 \\
              0 & 0 & 0 & \Delta_{\MS_{TN}} \\
              0 & 0 & 0 & 0 \\
            \end{array}
          \right) \anumber
\end{align}

\subsection{Angular momentum}
\label{relangmom}
The total angular momentum operator $\vec{\Jang}$ is once again defined by equation \eqref{Jang}. As usual we decompose the total moduli space into the centre of mass and relative moduli space. The vector fields generating the $SU(2)$ action on the Taub-NUT manifold  are denoted  $\xi^L_i$ in\mycite{GibbonsManton:CQDynBPSMon}. They are given by
\begin{align}
\xi^L_1 \ = \ & - \frac{\cos\phi}{\sin\theta} \frac{\partial}{\partial  \psi} + \sin\phi
\frac{\partial} {\partial \theta} + \frac{\cos\theta}{\sin\theta} \cos\phi \frac{\partial}{\partial \phi} \nonumber \\
\xi^L_2 \ = \ & - \frac{\sin\phi}{\sin\theta} \frac{\partial}{\partial \psi} - \cos\phi
\frac{\partial} { \partial \theta} + \frac{\cos\theta}{\sin\theta} \sin\phi\frac{\partial} {\partial \phi} \anumber \\
\xi^L_3 \ = \ & - \frac{\partial}{\partial \phi} \nonumber
\end{align}
and satisfy $\left[\xi^L_i, \xi^L_j\right] = \varepsilon_{ijk} \xi^L_k$. Again, the Lie derivatives with respect to these vector fields obey indeed the commutation relations \eqref{commLieadI} with the complex structures. The total angular momentum operator acts on $\overline{\alpha}_1$ and $\overline{\alpha}_2$ as
\begin{align}
\Jang_i(\overline{\alpha}_j) & \ = \ \half (\pauli_i)_{jk} \overline{\alpha}_k\eqip{,} \anumber
\end{align}
i.e.  $\{\overline{\alpha}_1, \overline{\alpha}_2\}$ form an angular momentum doublet.

\section{Outlook}

By applying supercharges to the bosonic scattering wavefunctions  $\Phi$ on the Taub-NUT manifold
found in section~\ref{scattstates} we generate a quartet of scattering states:
$\Phi$, $\overline{\partial} \Phi$, $\overline{\partial}_{\JCS} \Phi$, and $\overline{\partial} \, \overline{\partial}_{\JCS} \Phi = - \overline{\partial}_{\JCS} \overline{\partial} \Phi$ (or their analogues in the
spinorial description). Wedging these with the
a quartet of centre of mass states as outlined  in  section~\ref{sectexN=2} one obtains a 16-dimensional multiplet of supersymmetric monopole scattering states.  The total angular momentum
of these states can be computed using the general formula \eqref{Jang} and the
expressions for the angular momentum action on the centre-of-mass states and on the relative wavefunction, as discussed in sections~\ref{sectEx1Spin} and \ref{relangmom}.

It remains a challenge to interpret the resulting multiplets of scattering states
in terms of monopole-monopole (or dyon-dyon) scattering. One would like to be able
to compute spin-polarised differential scattering cross sections, where spins of the
in- and out-going monopoles  or dyons  are specified.  To do this in practice one
needs to relate the individual monopoles' spin degrees of freedom to the differential forms
on the  centre of mass and relative moduli space. As explained in  the opening
paragraphs of section~\ref{ansect}, this is only  possible in the asymptotic region of the
moduli space,  where the corresponding monopoles are well-separated.
The starting point for such a calculation
 would thus  be the decomposition  of the
asymptotic region of the total  moduli space $\MS_{1,1}$ \eqref{11mopomod} into  copies of the
constituent monopoles' moduli spaces, as discussed in \mycite{GauLowe:DyonsSDuality,LeeWeinbergYi:EMDSU3Mon}.   Using this decomposition one
needs to  decompose scattering states on  $\MS_{1,1} $ into  states which have definite values of the spin for the constituent monopoles.

An important motivation for carrying out a detailed study of spin-polarised monopole-monopole
scattering  cross sections comes from the electric-magnetic duality conjecture,
summarised in our Introduction. As mentioned there, we regard the current paper
as a stepping stone towards a similar analysis  in  $N=4$
supersymmetric  Yang-Mills-Higgs theory, where there is evidence for the
validity of  S-duality conjecture. We intend to carry out the analysis
of spin-polarised scattering   in that context.

\section*{Acknowledgements}
EJdV is grateful to Jos\'e Figueroa-O'Farrill for useful discussions on the quantisation and supersymmetries of the effective action.
BJS would like to thank Conor Houghton and Joost Slingerland for discussions at an early stage of this project.

\end{document}